\documentclass[tradiabstract,twocolumn]{aa} 

\usepackage{graphicx}
\usepackage{xcolor}

\usepackage{txfonts}
\usepackage{amsmath}	% Advanced maths commands
\usepackage{amssymb}	% Extra maths symbols
\usepackage{soul}       % for strikeout, can be removed after comments are treated
\usepackage[utf8]{inputenc}
\usepackage{array}

\usepackage{hyperref}
\hypersetup{colorlinks,allcolors=blue}

\bibpunct{(}{)}{;}{a}{}{,}
\usepackage{lscape}

\newcommand{\kms}{km~s$^{-1}$}

\def\apjl{ApJL}
\def\apjs{ApJ Sup.}

\def\aap{A\&A}

\begin{document}

   \title{Strong lensing in UNIONS: Toward a pipeline from discovery to modeling \thanks{The catalog of the 133 candidates and the modeling results are only available in electronic form at the CDS via anonymous ftp to cdsarc.u-strasbg.fr (130.79.128.5) or via \url{http://cdsweb.u-strasbg.fr/cgi-bin/qcat?J/A+A/}}}

   \author{E.~Savary\inst{\ref{epfl}}\and
           K.~Rojas\inst{\ref{epfl}}\and
           M.~Maus\inst{\ref{epfl}}\and
           B.~Cl\'ement\inst{\ref{epfl}}\and
           F.~Courbin\inst{\ref{epfl}}\and
           R.~Gavazzi \inst{\ref{iap},\ref{ioa}}\and
            J.~H.~H. Chan\inst{\ref{epfl}}\and
           C.~Lemon\inst{\ref{epfl}}\and
           G.~Vernardos\inst{\ref{epfl}}\and
           R.~Ca\~nameras\inst{\ref{mpa}}\and
           S.~Schuldt\inst{\ref{mpa},\ref{tum}}\and
           S.~H.~Suyu\inst{\ref{mpa},\ref{tum},\ref{iaaas}} \and
           J.-C.~Cuillandre\inst{\ref{Sarclay},\ref{observatoireparis}}\and
           S.~Fabbro\inst{\ref{canadadatacentre}}\and
           S.~Gwyn \inst{\ref{canadadatacentre}}\and
           M.~J.~Hudson \inst{\ref{waterloo}, \ref{waterloo2}, \ref{perimeter}}\and
           M.~Kilbinger\inst{\ref{Sarclay}}\and
           D.~Scott\inst{\ref{vancouver}}\and
           C.~Stone\inst{\ref{Queen}}}

   \institute{Institute of Physics, Laboratory of Astrophysics, Ecole Polytechnique 
F\'ed\'erale de Lausanne (EPFL), Observatoire de Sauverny, 1290 Versoix, 
Switzerland \label{epfl}
    \and
Institut d'Astrophysique de Paris, UMR7095 CNRS \& Sorbonne Universit\'e, 98bis Bd Arago, 75014 Paris, France \label{iap} 
    \and
Institute of Astronomy, University of Cambridge, Madingley Road, Cambridge CB30HA, UK\label{ioa}
    \and
Max-Planck-Institut f\"ur Astrophysik, Karl-Schwarzschild Str. 1, 85748 Garching, Germany\label{mpa}
    \and 
Technische Universit\"at M\"unchen, Physik-Department, James-Franck-Stra\ss{}e~1, 85748 Garching, Germany\label{tum}
\and
Institute of Astronomy and Astrophysics, Academia Sinica, 11F of ASMAB, No.1, Section 4, Roosevelt Road, Taipei 10617, Taiwan\label{iaaas}
\and
AIM, CEA, CNRS, Université Paris-Saclay, Université de Paris,F-91191 Gif-sur-Yvette, France \label{Sarclay}
\and
Observatoire de Paris, PSL Research University 61, avenue de l’Observatoire, F-75014 Paris, France\label{observatoireparis}
\and
Waterloo Centre for Astrophysics, University of Waterloo, 200, University Ave W, Waterloo, ON N2L 3G1 , Canada, \label{waterloo}
\and
Department of Physics and Astronomy, University of Waterloo, Waterloo, ON, N2L 3G1, Canada   \label{waterloo2}
\and
Perimeter Institute for Theoretical Physics, 31 Caroline St N, Waterloo, ON N2L 2Y5, Canada\label{perimeter}
\and
Canadian Astronomy Data Centre, Herzberg Astronomy and Astrophysics,5071 West Saanich Rd, Victoria, BC V9E 2E7,Canada \label{canadadatacentre}
\and 
Department of Physics and Astronomy, University of British Columbia, 6225 Agricultural Road, Vancouver, V6T 1Z1, Canada\label{vancouver}
\and
Queen's University, Dept. of Physics, Engineering Physics and Astronomy,Kingston, Canada\label{Queen}
}

  \abstract{We present a search for galaxy-scale strong gravitational lenses in the initial 2\,500 square degrees of the Canada-France Imaging Survey (CFIS). We designed a convolutional neural network (CNN) committee that we applied to a selection of 2\,344\,002 exquisite-seeing $r$-band images of color-selected luminous red galaxies (LRGs). Our classification uses a realistic training set where  the lensing galaxies and the lensed sources are both taken from real data, namely  the CFIS $r$-band images themselves and the  Hubble Space Telescope (HST). A total of 9460 candidates obtain a score above 0.5 with the CNN committee. After a visual inspection of the candidates, we find a total of 133 lens candidates, of which 104 are completely new. The set of false positives mainly contains ring, spiral, and merger galaxies, and to a lesser extent galaxies with nearby companions. We classify 32  of the lens candidates as secure lenses and 101 as maybe  lenses. For the 32 highest quality lenses, we also fit a singular isothermal ellipsoid mass profile with external shear along with an elliptical Sersic profile for the lens and source light. This automated modeling step provides distributions of properties for both sources and lenses that have Einstein radii in the range $0.5\arcsec<\theta_E<2.5\arcsec$. Finally, we introduce a new lens and/or source single-band deblending algorithm based on auto-encoder representation of our candidates. This is the first time an end-to-end lens-finding and modeling pipeline is assembled together, in view of future lens searches in a single band, as will be possible with Euclid.}

   
   %\authorrunning{Savary et al.}
   
   \maketitle
%
%-------------------------------------------------------------------

\section{Introduction}

\begin{figure*}[t!]
\centering
\includegraphics[width=18cm]{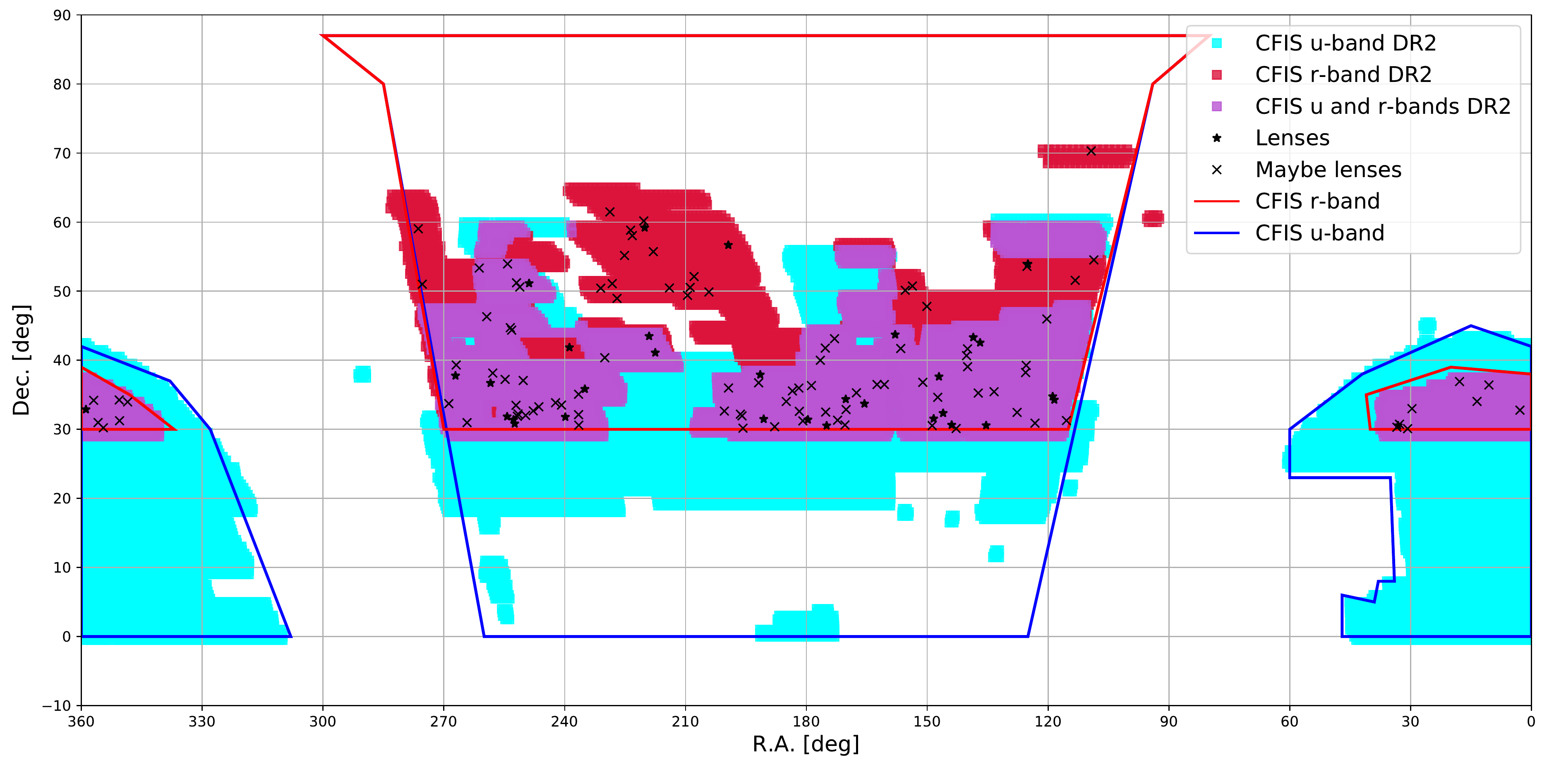}
\caption{Illustration of the planned and current CFIS footprint. The contours of the final CFIS-$r$ footprint and the  CFIS-$u$ footprint are shown in red and blue, respectively. Shown are the current areas covered in the $r$ band for DR2    (in red),   in  the $u$ band  (in blue),   and where  $r$-band and $u$-band data are available simultaneously (in violet). In this work the $u$ band is used, when available, for visual inspection purpose only. Also indicated is  the position of the 32 highest quality candidates (stars) and the 101 maybe lenses (crosses) obtained after the joint visual inspection of the CNN-committee candidates. Of these candidates 104 are new.}
\label{Fig: footprint3}
\end{figure*}
   
Strong gravitational lensing provides a unique astrophysical tool, via the formation of several distinct images of a high-redshift source. Depending on the source light profile and lensing mass distribution, multiple images of the source can appear as partial arcs or even complete arcs called Einstein rings. Such strongly lensed systems offer a vast range of astrophysical and cosmological applications, from the determination of cosmological parameters \citep{Suyu2017,Bonvin2017,Wong2020} to the study of galaxy mass profiles \citep{Koopmans2003,Sonnenfeld2015,Bellagamba2017} and halo substructure \citep{Mao1998,Dalal2002,Koopmans2005,Vegetti2009,Vegetti2010,Vegetti2012,Nierenberg2013,Vegetti2014,Hezaveh2016,Despali2016,Gilman2017,Vegetti2018,Chatterjee2018,Ritondale2019}. Observations of such lenses provide important calibrations for N-body cosmological simulations  \citep[e.g.,][]{Peirani2019,Mukherjee2021} and allow deeper higher resolution views of faint distant galaxies otherwise too faint to be studied \citep[e.g.,][]{Paraficz2018}.
However, due to the rarity of lens systems, many of these studies are limited by small sample sizes, prompting targeted lens searches by the community.

Lens searches can be divided into two broad classes: source-selected and lens-selected. The first  requires follow-up  of a known high-redshift source, in the hope of observing signs of strong lensing. Examples include early lens searches, such as the Cosmic Lens All-Sky Survey \citep[CLASS,][]{myers2003, browne2003} and the SDSS Quasar Lens Search  \citep[SQLS,][]{Oguri2006}. In contrast, lens-selected searches look for signs of a lensed high-redshift source in imaging or spectroscopy in known samples of massive galaxies. Some of the most well-studied lens systems come from the Sloan Lens ACS survey (SLACS) \citep{Bolton2006, Bolton2008}, the BOSS Emission Line Lens Survey \citep[BELLS,][]{Brownstein2012}, and the SLACS for The Masses Survey \citep[S4TM,][]{Shu2015}, which pre-selected high velocity-dispersion galaxies from SDSS, with signs of emission lines of a higher redshift source blended in the spectra. High-resolution HST imaging subsequently confirmed many of these systems as lenses. Lens searches are not limited to the visible domain. Similar efforts are done at longer wavelengths, particularly in the  submillimeter regimes, even if the search methods are then very different and mostly done by taking advantage of the magnification bias at the catalog level. Typical examples are searches in the Herschel Astrophysical Terahertz Large Area Survey \citep[][]{Bussmann2013, Wardlow2013, Nayyeri2016, Negrello2017}, the  South Pole Telescope Survey Data (SPT) \citep{Vieira2010,Vieira2013}, and the Planck all-sky survey \citep{Canameras2015}. With numerous ongoing and upcoming wide-field imaging surveys, such as \textit{Euclid} \citep[e.g.,][]{Laureijs2011, Amiaux2012},\textit{Roman} \citep{Spergel2015}, and the Rubin Observatory Legacy Survey of Space and Time (LSST) \citep[e.g.,][]{Ivezic2019}, lens samples can be built directly from imaging data. Previous imaging-only samples have come from visual inspection of HST images \citep{Faure2008,Pawase2014} or ground-based imaging with the help of citizen science \citep[e.g.,][]{Sonnenfeld2020}. With increasing depth and survey areas, visual searches alone become unsustainable and require automated techniques to condense the sample size. Early work included ring-finding algorithms, as was done with the SL2S sample \citep{Cabanac2007,Gavazzi2012} or model-aided search in HST \citep{Marshall2009} and HSC \citep{Chan2015}. More recently, machine-learning methods have been applied to ground-based surveys, including the Kilo Degree Survey \citep[KiDS, e.g.,][]{Petrillo2017,KIDS2019,Petrillo2019}; the Dark Energy Survey \citep[DES, e.g.,][]{DES2005,Jacobsa2019,Jacobs2019b}, and the Hyper Suprime-Cam SSP Survey \citep[HSC, e.g.,][]{HSC2018, Sonnenfeld2018,Wong2018,Sonnenfeld2019,Chan2020,Sonnenfeld2020,Canameras2020,Jaelani2020,Jaelani2021}.

The application of machine-learning techniques to lens searches encompasses a wide range of methods, from support vector machines to deep neural networks. However, in recent years convolutional neural networks (CNNs) have emerged out thanks to their well-tested reliability for image classification \citep{He2015}. In particular, the top five algorithms of the first strong gravitational lens finding challenge \citep{Metcalf2019} were mainly CNN-based. In this challenge CNNs were able to recover 50\%\ of the lenses. However, false positives, such as ring galaxies, spirals, mergers, or galaxies with companions, were a severe problem for CNNs trained with overly simplistic simulations. Using a training set that is as realistic and exhaustive as possible is thus crucial. The current number of known lenses in the CFIS footprint is very small, which  complicates the composition of a training set for machine-learning algorithms. For this reason, we still use simulations in this paper. However, recent lens searches conducted in the Dark Energy Spectroscopic Instrument Legacy Imaging Surveys' Data Release 8 have proven the possibility to train neural networks with a small number of lenses \citep{Huang2020, Huang2021}. We therefore hope to reuse our lens candidates to train the next versions of classifiers with real lenses only or a mix of simulations and real lenses. A typical approach to producing training sets for CNNs is to use entirely synthetic images, as in \citet{Jacobsa2019}, where the lens light, the lens mass, and the light profile of the source are analytical. Random images are then taken from real data and added to the simulated ``clean'' lenses to introduce instrumental effects and more realistic features, such as companions around the central galaxy. The main advantage of this approach is the ability to control the distribution of the lensing parameters. However, it may be difficult to reproduce all the complexity of real lenses with this method.

One approach to mitigating this consists in creating training sets that combine images of real foreground galaxies with simulated sources, as in \citet{Petrillo2017} and \citet{Pourrahmani2018}. In this  case the training set contains deflectors with more realistic light profiles. 
In the present work we go one step further toward more realistic simulations by also using  a real image for the background source, as in \citet{Canameras2020}. The main difference with \citet{Petrillo2017} and \citet{Pourrahmani2018} is that in our case only the lensing effect is simulated;  the shape and the light profile of both the deflector and the source are taken from real data.
In general, any classification performed by the CNNs is imperfect and candidates must be confirmed by follow-up observations. For efficient use of telescope time, the candidates are first visually inspected in order to remove the most obvious false positives. The timescale for this visual inspection must remain reasonable, in particular for future large-scale surveys. Therefore, it is crucial to keep a very low false positive rate. A common way to reduce the occurrence of false positives is to increase the proportion of the most common misleading objects in the negative training set \citep[e.g.,][]{Canameras2020}. However,  these must be taken as often as possible from real data. Simulating negative examples is not reliable enough, and using modified images from other surveys involves re-sampling and PSF mismatch, not to mention K-corrections and evolution effects due to redshift mismatch between surveys. We thus believe it is important to provide a catalog of false positives taken from our specific data. These false positives can be used to improve the training sets of future searches in CFIS.

In this paper we use CNNs to look for lensed galaxies in 2\,500 $\deg^2$ of the excellent-seeing $r$-band imaging of CFIS. This complements past and ongoing searches mostly carried out in the south. We first describe the data in Sect.~\ref{sec:data}, along with our machine-learning method and simulation pipeline in Sect.~\ref{sec:method}, and then carry out a visual inspection of the machine-classified objects in Sect.~\ref{sec:visual inspection} to remove false positives. In Sect.~\ref{sec:deblending} we present a single-band method based on auto-encoders that separates the lens and source light and enhances the contrast of each component. We additionally carry out automated mass modeling of the best candidates in Sect.~\ref{modeling}, and derive basic properties of the lens and source populations. Finally, in addition to our best candidates, we provide a catalog of false positives, which will be useful for future lens searches based on  neural networks. To our knowledge, it is the first time a full lens-finding and modeling pipeline has been presented  for single-band data. Although it specifically targets galaxies lensed by luminous red galaxies (LRG), we expect to make it more general when looking for lenses in the full CFIS footprint and for all types of lens galaxies.

 \begin{figure}[t!]
   \centering
   \includegraphics[width=9cm]{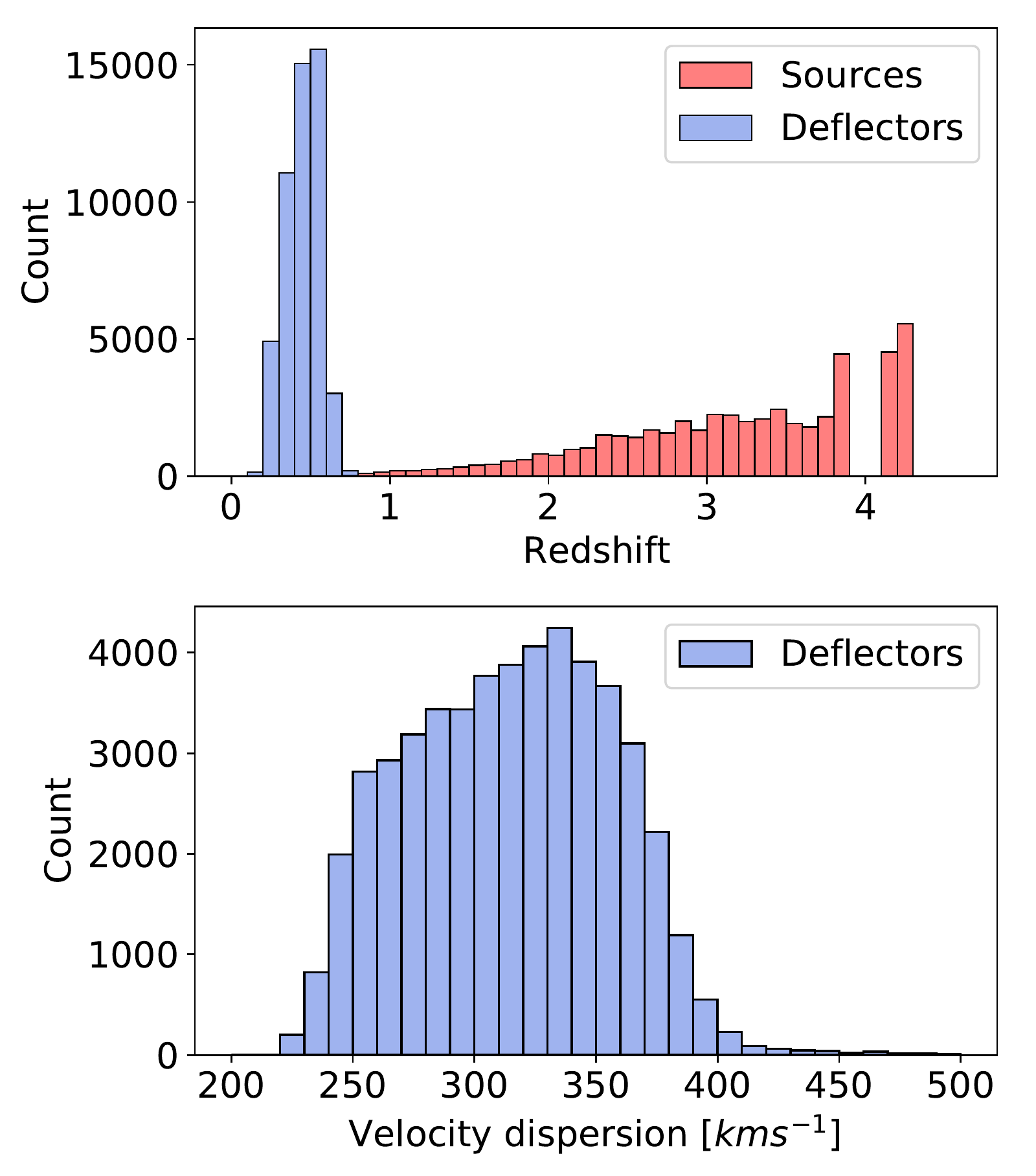}
      \caption{ Statistics of the deflectors and sources used in the lens simulations
 \textit{Upper panel:} Redshift distribution of the sources and deflectors used for the lens simulations. \textit{Lower panel:} Measured velocity dispersion distribution for the deflectors.} \label{Fig: statistic_spectro_selec}
\end{figure}

%--------------------------------------------------------------------
\section{Data}
\label{sec:data}

The Canada France Imaging Survey (CFIS) is an ongoing legacy survey using the Canada-France-Hawaii Telescope (CFHT), a $3.6$~m telescope at the summit of Mauna Kea in Hawaii. CFIS is a component of the multi-band Ultraviolet Near Infrared Optical Northern Survey (UNIONS). This effort will the provide the necessary ground-based optical counterparts of the forthcoming \textit{Euclid} space mission, along with stand-alone immediate scientific applications \citep[][Guinot et al. in prep.]{Ibata2017, Fantin2019}. When completed, the survey will have imaged 8\,000 $\deg^2$ of the northern sky in the $u$ band (CFIS-$u$) and 4\,800 $\deg^2$ in the $r$ band (CFIS-$r$).  The imaging data used in this work are from the CFIS Data Release 2 (DR2; see Fig.~\ref{Fig: footprint3}), covering around 2\,500 $\deg^2$. CFIS-$r$ has exquisite image quality with a median seeing of 0.6\arcsec, down to a depth of $24.1$ (point source, 10$\sigma$ {rms}). CFIS-$u$ has a median seeing of 0.8\arcsec\ to a depth of 23.6 (point source, 10$\sigma$ { rms}). Here we use the CFIS-$r$ footprint to search for lenses with CNNs based purely on high-resolution morphological information. Since not all $r$-rand images have $u$-band counterparts, the CNNs use only $r$-band information. However, CFIS-$u$ is used, when available, to refine the sample of candidates found through the CNNs by visual inspection. For each target, we have also produced models of the point spread function (PSF) and its spatial variations across co-added images which were   reduced, processed, and calibrated at the Canadian Astronomical Data Centre using an improved version of the \texttt{MegaPipe pipeline} \citep{Gwyn2008}. For each lens candidate we exploited the model PSF obtained with \texttt{PSFEx} \citep{Bertin2011} to produce an image of the local PSF oversampled by a factor of 2. Weight images along with other data quality diagnostics are also produced for each candidate in each of the available bands. Details of the spatial variations of the PSF may not always be well accounted for in such a model performed on stacked data, involving largely dithered exposures. This is, however, not a major issue for the strong lens modeling applications in this work.

Our goal is to provide an automated pipeline to find, deblend, and model high-quality galaxy-scale lenses. Our sample is by construction lens-selected, meaning that we look for lensed systems among a large sample of pre-selected LRG. These objects are bright and massive, and are therefore expected to have the largest possible lensing cross section. \citep{Turner1984}. 

\subsection{Data selection for the lens search}
\label{sample}

A reliable color selection of LRGs is not possible with CFIS data alone. Even with color information, it is necessary   to account for the fact that LRGs acting as lenses have colors biased toward the blue with respect to LRGs without lensing features: lensing LRGs are blended with the lensed image of a  background galaxy (which is often
blue). Fortunately, the CFIS footprint is entirely included in the first part of Pan-STARRS1 (PS1), hence we used PS1 to carry out our color selection, thus accounting for the blue bias. This has already been implemented by \citet{Canameras2020} and consists of a color cut in the PS1 3$\pi$ catalog, broadly matching the aperture magnitudes and colors of 90 000 Pan-STARRS simulations of lensing systems. This photometric selection is very large, since it was designed to include 96\% of the mock lens, and thus hopefully all LRG lens galaxies. In return,  however, it may  contain a large number of interlopers such as spirals and rings due to the Pan-STARRS data quality. After a cross-match with this catalog, we obtained 2\,344\,002 images to carry out our lens search.

\subsection{Data selection for the simulated training set}
\label{trainingset}

Our simulation set is constructed from real data (i.e., a deflector from CFIS imaging data and a background source from HST images), as described in Sect.~\ref{sec:simu}. The image stamp size in this work is 8.17\arcsec\ per side corresponding to 44 pixels.

The selection of deflectors is taken directly from \citet{Canameras2020}. This corresponds to a subsample of the LRG spectroscopic sample \citep{Eisenstein2001}, which uses color-magnitude cuts to select intrinsically red and luminous galaxies. They have SDSS spectra, and thus also velocity dispersion ($\sigma_\star$) and redshift ($z$) estimates. In Fig.~\ref{Fig: statistic_spectro_selec} we summarize the spectral properties of our selection, which spans the ranges $200 < \sigma_\star < 500$ \kms\ and $0.1 < z < 0.7$. After a cross-match with the whole CFIS-$r$ catalog from DR2, we obtained 624\,170 LRG images, which form the basis of our training set.

The background galaxies were taken from the sample of \citet{Canameras2020}. We used galaxy morphologies from HST/ACS F814W images and converted to r band using HSC ultra-deep stacked images. The original stamps have a size of 10\arcsec\ per side and the same pixel size as the HST/ACS F814W image (i.e., 0.03\arcsec). Since the PSF of the HST images is much sharper than the CFIS PSF, we neglected its effect during the simulation process and we did not attempt to deconvolve the HST images from their PSF. All sources are included in the COSMOS2015 photometric catalog \citep{Laigle_2016} and in the Galaxy Zoo catalog \citep{willett2017}. The redshift information for our sources, when available, was obtained from public spectroscopic catalogs \citep{Lilly2007,Comparat2015,Silverman2015,Lefevre2015,Tasca2017,Hasinger2018}. When no spectroscopic redshift was available, the best photometric redshift estimate from \citet{Laigle_2016} was used. Then with all this information combined we obtained high-resolution \textit{r}-band images of unlensed sources with known redshift. 
These selected foreground LRGs and background HST sources provided the basis for building our training set for the CNN search, as described in Sect.~\ref{sec:simu}

\section{Method}
\label{sec:method}

One way to address the problem of lens detection is to consider it as a binary image classification task where the positive class members are the lenses and the negative class members are the galaxies without lensing features. CNNs are especially suited to this task as they are able to detect local correlations of two-dimensional features in images \citep{Lecun2015}.  The convolutional layers of CNNs can be understood as a set of kernels that act as specific feature detectors.

\subsection{Classifier}

In this work we use a recent class of CNNs called EfficientNets \citep{tan2020efficientnet}. They outperform the most common CNN architectures on the classification of images from different standard  data sets while using a smaller number of parameters \citep[see Fig.~1 of][]{tan2020efficientnet}. This is achieved by scaling uniformly the depth, width, and resolution as a function of the available computing resources.

In our case we did not scale the models ourselves, but   used the models already implemented in the \texttt{Keras} application programming interface (API) \citep{chollet2015keras}. It includes eight versions of EfficientNet, named B0 to B7, depending on the number of free parameters involved. The dimensions of our images, 44 pixels per side, are much smaller than the dimensions of the images of the standard machine-learning data sets used in \citet{tan2020efficientnet}. Therefore, we used the B0 architecture, which contains the smallest number of parameters of the eight models. The B0 architecture, from the Keras API is  pre-trained on ImageNet data \citep{Deng}. We took advantage of this pre-training by reusing the parameters of the trained model instead of initializing them randomly, which allowed us to speed up the training. However, since the dimensions of our images are different, we adapted the size of the first layer and the last fully connected layers and randomly initialized the parameters of these layers. The classification was performed using the so-called ensemble-averaging method. This consists of separately training models with the same architecture but different subsets of the training set and combining their results in order to reduce the variance of the predictions. In the following we call the set of models a ``committee'' and each individual model a ``member'' of the committee. Here, we use three separate instances of EfficientNet B0 as our committee members. We also tested versions of the committee with more instances. This did not lead to a significant improvement in the quality of the classification.

\subsection{Design of the training set}
\label{sec:simu}

Since we performed a binary classification we needed to build a training set containing images with either positive or negative labels. The negative examples were drawn randomly from the lens search sample described in Sect.~\ref{sample}. The negative examples may contain a few real lenses, but we expected that this would have only a marginal effect on the performance of the network since the prevalence of gravitational lenses is very low. 
Creating a set of positives examples required more preparation. Not enough lenses have been confirmed in the CFIS footprint to build a training set that   spans the full diversity of lens systems. We therefore generated a set of simulated lenses using the pipeline described in \citet{Schuldt2020holismokes}. We present below the most important steps of this process. 

We first selected an LRG image from the deflector catalog constructed in Sect.~\ref{trainingset} and assigned a mass profile to the selected galaxy assuming a simple parametric model, the singular isothermal ellipsoid (SIE) \citep{Kassiola1993,Kormann1994}. This mass model has five free parameters: the Einstein radius, the coordinates of the lens center, the ellipticity (or axis ratio), and the position angle (PA). The lens center coordinates were fixed to the center of the deflector image, whereas the values of the axis ratio and the position angle were derived from the second moment of the lens light profile. In this model we note that the  ellipticity and PA are assumed to be the same for the light and mass distribution.\\A source was then randomly selected from the source catalog. Knowing both the deflector redshift and velocity dispersion, we computed the Einstein radius and checked, given the redshift of the source, that it fell in the range $0.8\arcsec < R_{\rm E} < 3.0\arcsec$. The lower limit was chosen to prevent lensing features becoming blended with the deflector light. If the Einstein radius was outside the given range, we randomly selected another source from the catalog. We repeated this until a matching source was found, otherwise after 100 iterations we increased the velocity dispersion (and hence the Einstein radius) of the deflector by 50\%\ and   repeated the process. The goal was to obtain a sufficient number of simulations. If no match was found after increasing the velocity dispersion, we discarded this deflector from our catalog. Since this boost involved  only a few objects with small velocity dispersion at the lower end of the distribution shown in Fig. ~\ref{Fig: statistic_spectro_selec}, we did not expect this to introduce a morphological bias in our training set.

\begin{figure}[t!]
\centering
\includegraphics[width=\hsize]{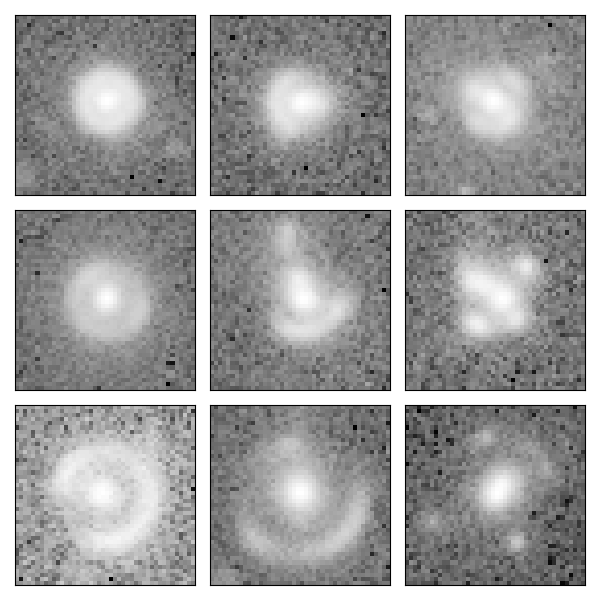}
\caption{Examples of simulated galaxy-scale systems of the CFIS survey. Each stamp is the sum of a real CFIS $r$-band image of a galaxy, to which a lensed HST galaxy convolved with the CFIS PSF is added. Each stamp is 8.17\arcsec\ per side, and the pixel size is 0.18\arcsec.}
\label{mosaic_simu}
\end{figure}

In the next step the position of the source was chosen randomly in the source plane. However, we imposed a total magnification constraint, $\mu \ge 2 $. This limit corresponds to the minimum magnification threshold to produce multiple images. Choosing a higher limit gives more striking lensing features, but also artificially increases the proportion of Einstein rings among the simulations. This may bias the classifier toward this class of objects or even lead to more false positives, such as ring galaxies. Only the positions resulting in a magnification $\mu \ge 2 $ were considered, hence constraining the source to be within or close to the caustics and resulting in multiple images. Once the source position was chosen, we computed a high-resolution image of the lensed source using the \texttt{GLEE} software package \citep{SuyuHalkola2010,Suyu2012}.

As a final step, the CFIS PSF was re-sampled to the HST pixel size and the image of the lensed source was convolved with this re-sampled PSF. The result of the convolution was then down-sampled back to the pixel size of CFIS and added to the deflector image. Our simulations are therefore a hybrid between simulations and real data (i.e., built from the CFIS data themselves for the lens and from deep HST images for the source).

For some of the simulations produced with this method the lensing features are too faint or too heavily blended with the deflector light. Including images with indistinguishable lensing features may increase the false positive rate. Therefore, we used only the simulations for which the sum of the brightness of all pixels of the lensed source was at least 20 times the mean {rms} value of the sky noise measured in the four corners of the deflector image. We then proceeded to a rough visual inspection of all simulations above this threshold to remove images with lensing features blended in the deflector light. This resulted in 10\,600 accepted lens simulations, of which we show a few examples in Fig.~\ref{mosaic_simu}. This number is relatively small. However, since the precision and recall on the validation set are close to perfect, we do not expect that increasing the size of the training set would have a significant impact on the performance measured on validation data.

\subsection{Pre-processing and training}
Before being passed to the CNNs, all images were normalized so that the full dynamical range lies between 0 and 1. We also applied a logarithmic stretch in order to enhance the contrast of the lensing features. After this pre-processing, the data were separated into three different sets: (1) the training set (80 \% of the training data); (2) the test set (10 \%), and (3) the validation set (10\%). The validation set is used both to monitor the training process and to define the conditions to end the training, whereas the test set is used only at the end of the training to evaluate the performance of the committee.

Our images were only available in the $r$ band, but the EfficientNet architectures from the Keras API are built to handle three-band images. Hence, we transformed our single-band images into a three-band data cube by duplicating three times the images before passing them to the network. In doing so, we were able to use the pre-trained version of the network, allowing us to shorten the training time. We trained the three members of the committee independently (each member being an instance of the EfficientNet B0 architecture), each with a different subset of the training set. The subsets were constructed using a different fraction of lenses; the fraction was  drawn randomly in the range 0.2 to 0.5 to mitigate the tendency of the network to learn the fraction of positive examples seen in the training set. The instance trained with the lower fraction will be less optimistic and find fewer lenses, but will reach a lower false positive rate. However, the fraction of lenses is set to 0.5 in the validation and test sets.

To train the CNNs we performed a mini-batch stochastic gradient descent using binary cross-entropy as a loss function and an Adam  optimization ~\citep{Kingma2014}. The batches contained 128 images picked from the data set and flipped randomly along the $x$- and $y$-axes using the Data Augmentation method from Keras. Overfitting is one common pitfall encountered during the training of machine-learning algorithms. It occurs when the algorithm learns the specifics of the training data, and thus is not able to generalize on new data. When overfitting occurs, the classification error on the training data becomes very small, while the error on previously unseen data starts to grow. Since the batches contain  different images each time, the data augmentation procedure allows us to artificially increase the size of the training set and limit the risk of overfitting. Since our network was previously trained on ImageNet data, we only needed to fine-tune the parameters. Therefore we started directly with a relatively low learning rate of $10^{-4}$.

The maximum number of epochs, which is  the number of times the machine-learning algorithm is allowed to see the entire training data set, was fixed at 200. However, in order to optimize the training time, we used the early stopping procedure from the \textit{Keras} API, which interrupts the training before reaching the maximum number of epochs if the validation loss is no longer improving. More precisely, the training stops if the validation loss reaches a plateau, or if it increases during ten consecutive epochs, or if neither of the  other two conditions is met after 200 epochs. Using early stopping with a validation set allows us to interrupt the training if the classifiers start to overfit the training data.
At the end of the training, the weights and biases are restored to the epoch achieving the smallest validation loss. The three members of the committee are combined such that the final output corresponds to the mean of all the outputs assigned by the three independent networks. 

\begin{figure}
  \centering
  \includegraphics[width=\linewidth]{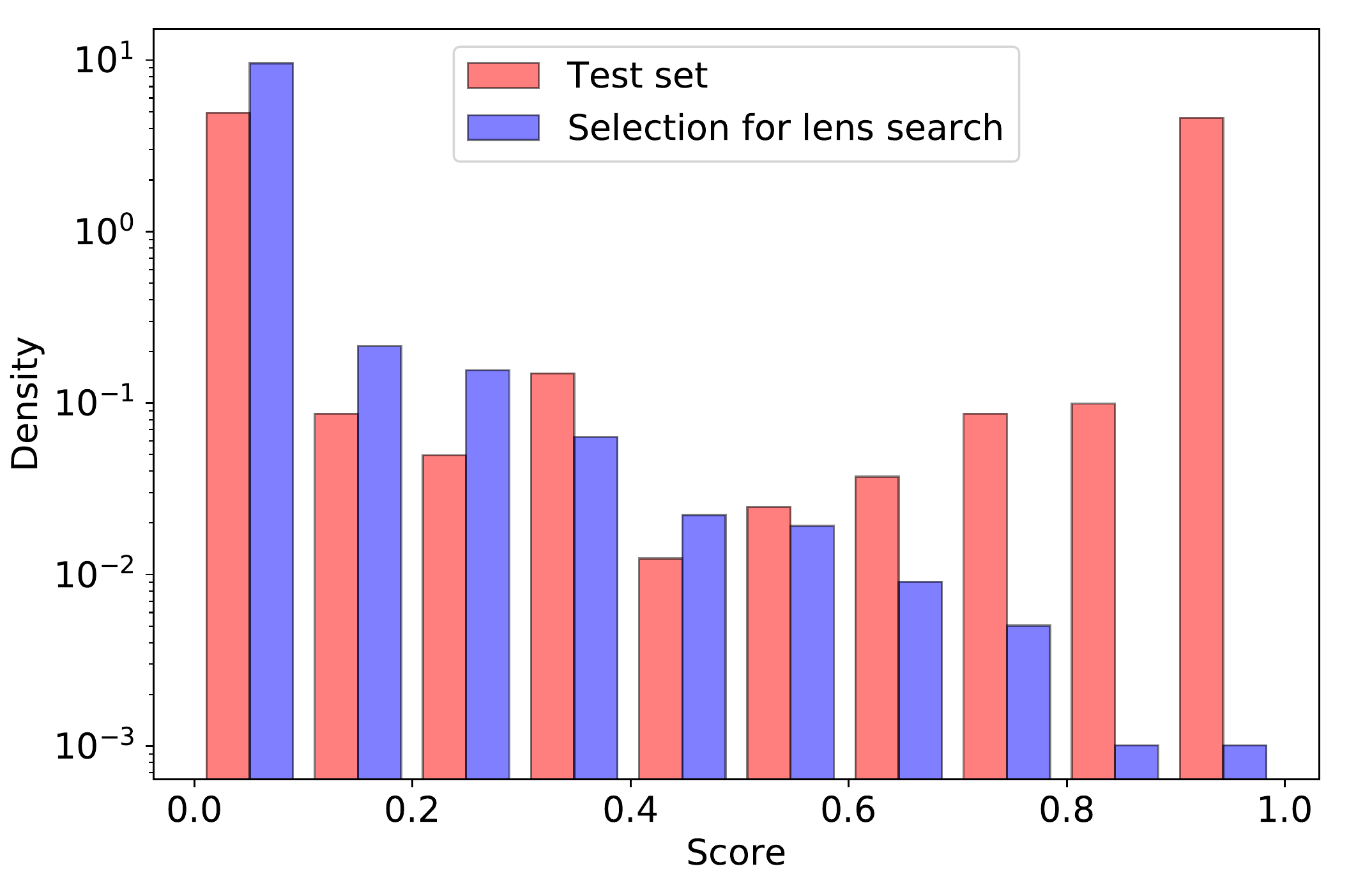}
  \caption{Probability density distribution of the score assigned by the committee for the test set (red), containing 50\% of lens simulations, and 50\% of galaxies without lensing features taken from the lens search sample. In blue are shown the scores predicted by the network on the data set containing real images described in Sect.~\ref{sample}. In the lens search sample the proportion of lenses is by definition unknown.}
  \label{Fig: distribution_score}
\end{figure}
   
\subsection{Candidate detection and performance of the classifier}
\label{subsec:performance classifier}

Convolutional neural networks are generally not invariant under rotation and the final output of the committee, hereafter referred to as the score, can change significantly for the same image if this image is rotated in different ways. In some extreme cases, an image with a high score can even fall under the selection threshold after a rotation due to statistical fluctuations. In order to mitigate this effect, we rotated and flipped all images in seven different ways:  three rotations of 90, 180, and 270 degrees, and flips along the $x$-axis of all rotations including the flip of the original unrotated image. We then considered the mean of the scores given by the committee in all directions as the final score.
The final scores of the committee range from 0 to 1. If the classifier were     ideal, we would expect the scores of galaxies without lensing features to be 0 and the scores of the lenses to be 1. Figure \ref{Fig: distribution_score} shows the distribution of scores assigned by the committee on all images from the lens search sample and for the test set. The test set distribution is not bi-modal, indicating that our classifier is not perfect. In the case of the lens search sample the distribution has only one peak centered on zero, and decreases exponentially afterward. This can be explained partially with the very low prevalence of lenses in the lens search sample and by the fact that the committee may less easily  identify lenses than simulations, as explained below. 
The performance of the network is evaluated using two metrics on the test set: Precision (P) and recall (R). The precision or purity indicates the fraction of true lenses among all images labeled as lenses, whereas the recall, also called ``true positive rate'' or ``completeness,'' gives the fraction of true lenses recovered by the committee among all the true lenses of the training set. They are defined as 
\begin{equation}
\label{Eq:precision}
    P=\frac{TP}{TP+FP},
\end{equation}
\begin{equation}
\label{Eq:recall}
    R=\frac{TP}{TP+FN},
\end{equation}
where $TP$, $FP$, and $FN$ are the number of true positives, false positives, and false negatives, respectively.

\begin{figure}
   \centering
   \includegraphics[width=8.9cm]{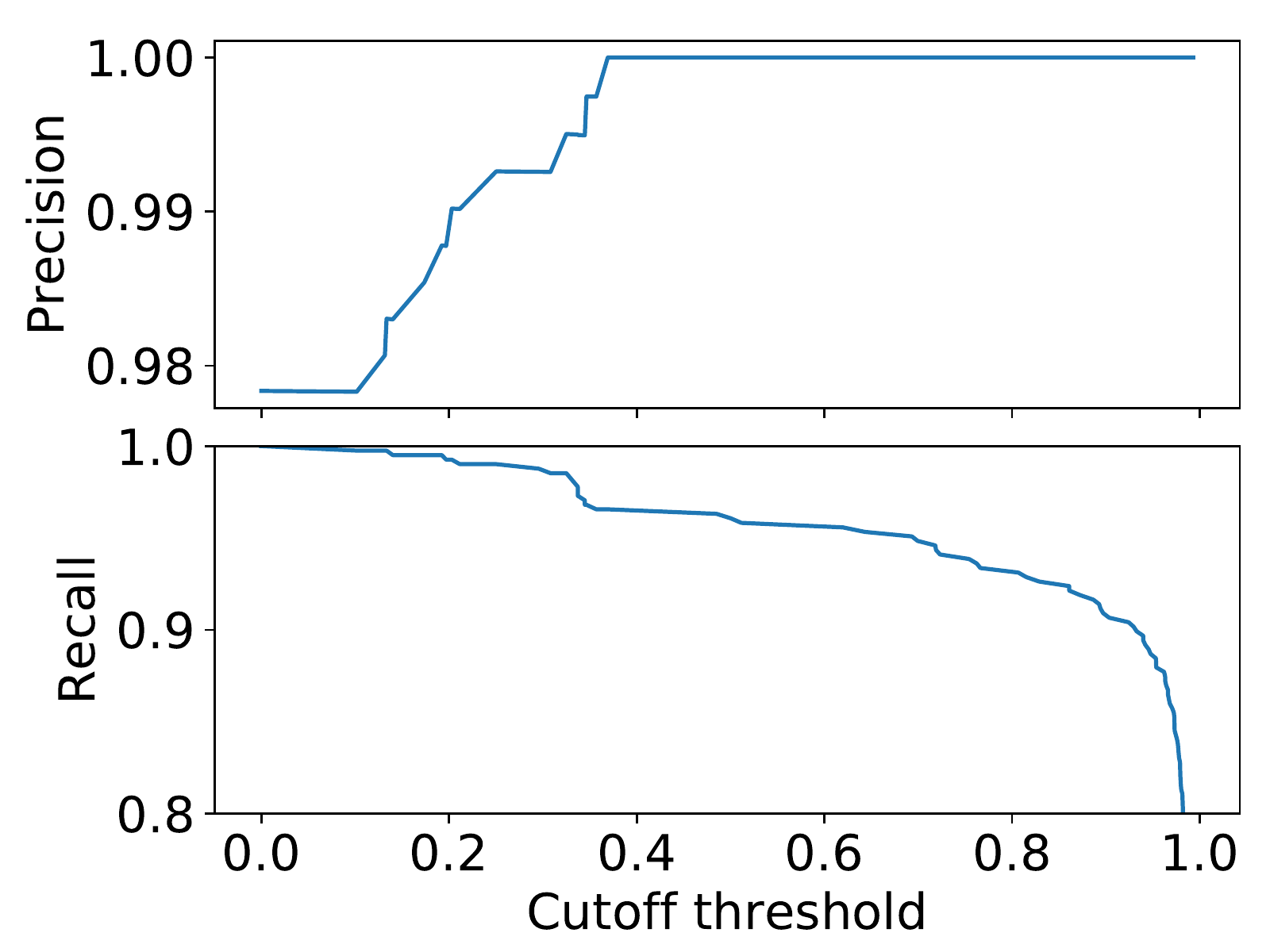}
      \caption{Precision and recall as  functions of the cutoff threshold in CNN scores applied to a test set composed of 1\,060 galaxies without lensing features taken from our selection for the lens search and 1\,060 lenses taken from our simulation set.}
         \label{Fig: precision_recall}
   \end{figure}

\begin{figure}
   \centering
   \includegraphics[width=9.2cm]{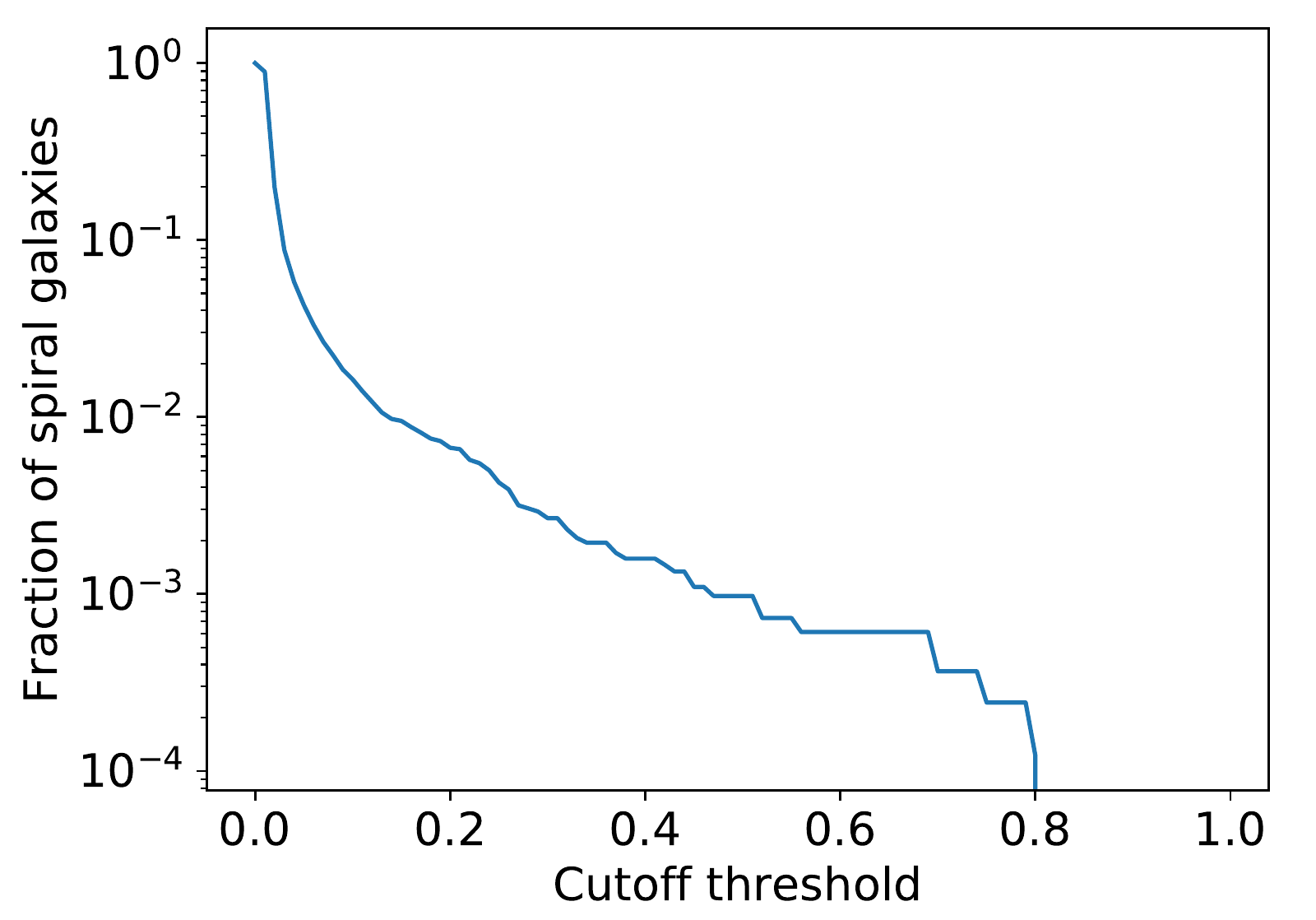}
      \caption{Fraction of CFIS $r$-band images of 8\,200 Galaxy Zoo spiral \citep{willett2017} galaxies mislabeled as lenses as a function of the cutoff threshold on the CNN-committee score.}
         \label{Fig: spiral fraction}
   \end{figure}

The number of true positives, true negatives, false positives, and false negatives from Eqs.~\ref{Eq:precision} and ~\ref{Eq:recall} depend on the score we chose as a cutoff threshold (i.e., the score above which the images are considered to be lens candidates).
We show in Fig.~\ref{Fig: precision_recall} the precision and recall values for all cutoff thresholds between 0 and 1. Choosing a high cutoff threshold increases the precision as the number of false positives decreases, but lowers the recall since fewer true positives are included. In Fig.~\ref{Fig: precision_recall} we observe that the precision and recall on the test set stay fairly high, independently of the cutoff threshold. 
Since a large part of the contaminants in CNN-based lens searches are spiral galaxies, we estimated the proportion of spiral false positives for each cutoff threshold. Therefore, we evaluated 8\,200 CFIS images of spiral galaxies with our trained CNNs. All of these spirals were taken from the Galaxy Zoo catalog \citep{willett2017}. The fraction of spirals mislabeled as lenses as a function of the cutoff threshold is presented in Fig.~\ref{Fig: spiral fraction} and shows that the contamination rate falls below $0.001$ for any CNN score above 0.5. For scores higher than 0.8 there are no spiral false positives.

Taking into account both the precision and recall curves, we chose a cutoff threshold of 0.5 for our lens search in the real CFIS data. This results in a precision of 1 and a recall of 0.96 on the test set. However, these results must be interpreted with caution, since the committee may have learned to recognize the simulations, and the performance may decrease on real data. In addition, it should be kept in mind that the real occurrence rate of strong lensing events is very low. The probability for an image classified as a candidate to be a real lens ($P(L|C)$) can then be deduced from the Bayes rule to take into account the occurrence rate ($P(L)$) of lenses in the data set as 
\begin{equation}
    P(L|C)=\frac{P(L)P(C|L) }{P(L) P(C|L)+(1-P(L))P(C|NL)},
\end{equation}
where $P(C|L)$ and $P(C|NL)$ are the precision and false discovery rate obtained on the testing set, respectively. If $P(L)$ is very low, $P(L|C)$ will stay low even when the precision is very high. This effect, called the  base rate fallacy, is known to limit the performance of intrusion detection algorithms (e.g., \citealt{Axelsson2000}) and may be non-negligible for lens detection.

\section{Visual inspection of the candidates}
\label{sec:visual inspection}

Of the 2\,344\,000 images selected for our lens search, 9\,460 obtained a CNN score higher or equal to 0.5. All these candidates were visually inspected independently by the six authors of this paper (E.S., B.C., K.R., F.C., J.C., and G.V.). We separated the candidates following the score attributed by the committee into five CNN-score bins of 0.1 in size and inspected them separately. This separation may introduce some biases in the visual inspection, meaning that the users may give more optimistic grades for objects in the bins corresponding to the highest score. In order have a consistent visual inspection between the users and the bins, we defined common guidelines and designed a quick and flexible {\tt python}  tool\footnote{\url{https://github.com/esavary/Visualisation-tool}} to display the images in two different ways: a {\tt mosaic} tool to display a large number of images simultaneously  and a {\tt single-object} display tool to review each object in detail. Both tools can handle single-band and color images.

For each score bin, we first performed a crude pre-selection of potential lenses using the {\tt mosaic} tool. This tool displays a grid of $10\times10$ cutouts so that users can quickly flag any of the systems. The images are arranged randomly in the mosaic for each user, hence minimizing any biases depending on the order and positions in which the images are shown, for example  due to varying attention levels of users through the inspection process. We also flagged misclassified ring galaxies in this pre-selection in order to improve our set of negative examples for future lens searches.

At the end of this pre-selection, we obtained a list of possible lens candidates and ring galaxies. This sample unavoidably contained other objects, such as spirals and interacting galaxies that resemble lenses, which were inspected in more detail in a second step (see below). In other words, this first step rejects any object that can be immediately demoted to a false positive and keeps the rest. We then considered the union of all objects selected by at least one of the six users (4\,626 images). 

For all images selected during the mosaic inspection, we   proceeded with a more detailed inspection using our {\tt single-object} tool. This tool displays one single stamp at a time, but offers a dynamic contrast control, more classification options, and a direct link to the Legacy Survey  \citep[LS;][]{Legacy} cutouts, when available. The LS data are shallower than the CFIS data, but have color information. We note that the LS color images are not displayed systematically during the classification process to avoid the users' decisions being driven by the color information. They are rather used to support the $r$-band CFIS classification when a candidate requires further data to make a decision. In this step we classify the images into four categories:  secure lenses (SLs), maybe lenses (MLs), single arcs (SAs), and non-lenses (NLs). 

Our single-object classification follows the same guidelines as in \citet{Rojas2021}, where the SL category includes images displaying obvious signatures of strong-lensing features, such as multiple images or arcs with counter images. The second category, ML, corresponds to images that exhibit structures compatible with lensing, but that would require further investigation with lens models or follow-up with higher resolution  data, higher signal-to-noise data, or  spectroscopy. When single arcs with small distortions and no counter-images are seen, we use the category SA. Naturally, the SA category contains fortuitous object alignments or galaxies with a curved shape that may not be due to lensing. All the images that do not belong to the first three categories are labeled   NLs. The NL objects can be subclassified into three subcategories: rings, spirals, and mergers. However, these subcategories are used by the users only if they are very confident about their classifications. This allows us to obtain a catalog of the most common false positives as a by-product of this search. These samples are valuable for future searches with CNNs being trained against false positives that can be reliably identified. At this stage, however, the size of this false positive sample remains too small to retrain the CNNs used in this work. 

We obtained a total of 1\,423 images selected as MLs or SLs by at least one user (out  of the six users). However, the agreement between the potential lens candidates selected by the different users remained low after the two-step visual inspection. More details about the classification are  given in Appendix  \ref{Sec:appendix visual inspection}. In order to obtain an agreement, we reinspected all images selected by at least one user in a joint visual inspection session. In doing so, we considered all 1\,423 objects that were   identified as SL or ML by at least one person and chose one unique grade for each object after a discussion among all the  users. In this last part, we also showed the image of the candidates in the $u$ band when available, and used only the categories SL, ML, and NL to further clean the sample. It should be noted that this process is very selective as we require the agreement of all users to grade an image as SL or ML.

After the joint inspection we obtained 32 objects classified as  SLs and 101 MLs, which are shown in Fig.~\ref{Fig: deblending lens 1} and Fig.~\ref{Fig:ML mosaic}, respectively. These represent 1.4 \% of all the candidates selected by the committee of CNNs. After a cross-match with Vizier \citep{vizier}, Simbad \citep{simbad}, the Master Lens database, and the catalogs of various lens-finding papers with candidates or confirmed lenses included in the CFIS footprint \citep[e.g.,][]{Canameras2020,Chan2020,Jaelani2020,Huang2021,Talbot2021}, we obtained 15 new SL and 89 new ML candidates. The Table of results of the cross-match and the coordinates of the candidates is available at the CDS. With our visual inspection, we also obtained three catalogs with 238 mergers, 361 ring galaxies, and 950 spiral galaxies identified by at least one user, which can be used to expand our negative sample for future searches. Examples of each category are shown in Fig.~\ref{Fig: FP mosaic}. Because we consider the unions of all votes to include a candidate in our false positive list, there is a small an overlap between the three catalogs.  Among all images labeled as ring or spiral by any user, 11 were finally classified as ML or SL after the joint inspection step. We therefore removed them from the final spiral and ring catalog. The relatively small number of ML and L candidates in our final catalog in comparison with the number of rings, spirals, and mergers can partially be explained by the visual inspection method: we require a unanimous decision of all users during the joint inspection to include objects in the ML or L, whereas only one vote is sufficient for an image to be included in the false positive catalogs. 

\section{Lens--source deblending with auto-encoders}
\label{sec:deblending}

Independently of the lens search itself, it is desirable to provide reliable deblending of the lens and source light of the candidates without relying on a lens model. First, deblending reduces the dynamical range of the data and allows  faint structures to be seen more clearly either in the lens or in the source. Second, it allows the remeasurement of  clean photometry of the lens and source for future photometric redshifts estimates when color information become available (e.g., from public release of other surveys). Finally, it can be used to initialize lens model parameters when implementing composite profiles with both stellar and dark mass.

\citet{Rojas2021} propose a method for deblending lens candidates based on the {\tt scarlet}\footnote{\url{https://github.com/pmelchior/scarlet}} \citep{Melchior2018} and {\tt MuSCADeT} \citep{Joseph2016} algorithms. However, this method is not directly applicable to our case since it requires color images. Therefore, we present here a fully data-driven alternative approach based on a class of neural networks called auto-encoders, with the goal of deblending our 32 SL candidates.
In general, neural networks find a mapping, $Y = f (X)$, between the inputs $X$ and the labels $Y$. In the case of auto-encoders the labels are the inputs themselves. In other words, the mapping made by the auto-encoder rather writes as $X = f (X)$. Auto-encoders can be decomposed into two symmetrical parts: the encoder and the decoder.
If the dimensions of the layers decrease from the two ends of the auto-encoder to the central layer, the network is able to learn a simplified representation of the original input. Thus, auto-encoders may be used for data compression, feature learning, dimensionality reduction, and denoising.
The architectures derived from auto-encoders, like variational
auto-encoders, can also be used as generative models, to then generate realistic images of galaxies \citep{lanusse2020deep}.

The scheme of the auto-encoder we used for the deblending is presented in Fig.~\ref{Fig: auto-encoder architecture}. The input of the network is the image of the lensed system. Unlike traditional auto-encoders, the decoder is split into two parallel parts. The first part extracts the lensed source, whereas the second extracts the deflector image. The dimension of the inner dense layers correspond to the flattened dimension of the last convolutional layer of the encoder and the first of the decoder part. In the end, we obtained  three different outputs: the lensed source, the deflector and the lens system reconstructed by the auto-encoder. The reconstructed lens system was obtained by summing the lensed source and deflector images derived with the two different parts of the decoder.

\subsection{Training process}
We trained the auto-encoder using 10\,000 simulations of lenses taken from the sample described in Sect.~\ref{sec:simu} and 5\,000 images of LRGs from the spectroscopic LRG selection detailed in Sect.~\ref{sec:data}. Before the training all images were normalized between 0 and 1. We also set aside 20\%\ of them to constitute the validation set, with the rest as the training set. For each image, we used the following for the ground truth: the image of the lens system itself, and the lensed source and deflector images obtained in the final step of the simulation pipeline before the two images were combined. In the case of LRG-only images, we define the lensed source image as an array of zeros and the deflector image as the image of the LRG. The loss function takes into account the three different outputs and gives more weight to the part containing the lensed source and deflector terms in order to put emphasis on the accurate deblending of the images. It is defined as 
\begin{equation}
   L= 0.4 \times L_{\textrm{deflector}} + 0.4 \times L_{\textrm{source}} + 0.2 \times L_{\textrm{combined}},
\end{equation}
where $L_{\textrm{deflector}}$, $L_{\textrm{source}}$, and $L_{\textrm{combined}}$ are respectively the binary cross-entropy losses computed between the true deflector image and the deblended deflector, between the lensed source image and the deblended lensed source, and between the lens system image and the combination of the deblended deflector and the lensed source. We also tested a combination of mean squared error losses instead of binary cross-entropy losses. However, the version trained with mean squared error failed to  correctly restore the shape of the lensed source. In most of the case the lensing features appeared incomplete or distorted, or were absent.

We set the maximum number of epochs to 200 and use early stopping to avoid overfitting.

\subsection{Secure lens deblending}

\begin{figure*}[htbp]
 %\centering
    
    \includegraphics[width=\linewidth]{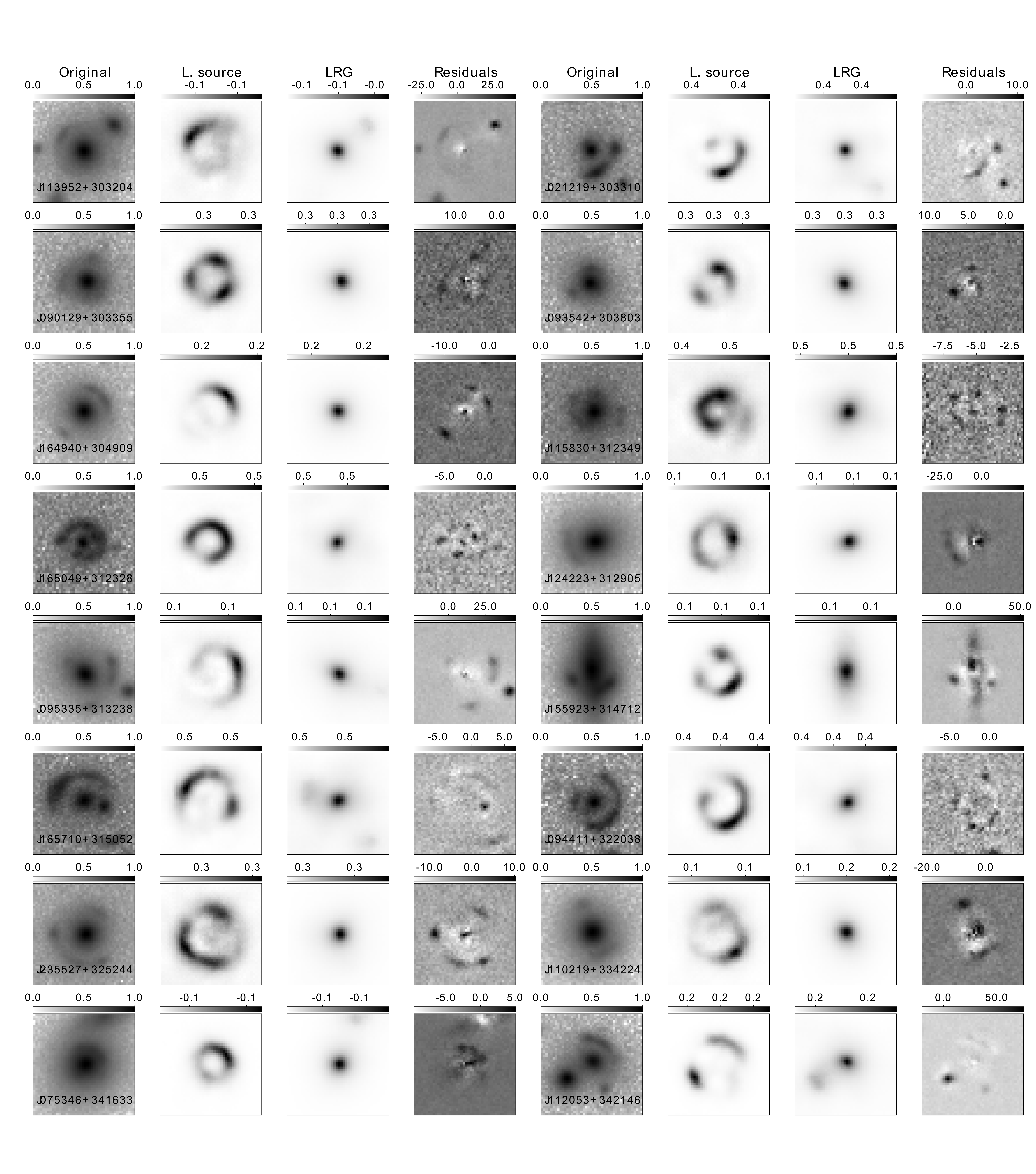}
    
    \caption{Results of the auto-encoder deblending for our 32 SL candidates. Two objects are displayed in each row. For each object, we show the original image displayed using a asinh grayscale, and  the lensed source and lens light   deblended using the auto-encoder and the scaled residuals.}
    \label{Fig: deblending lens 1}
\end{figure*}

We present in Fig.~\ref{Fig: deblending lens 1} the result of the auto-encoder deblending for the 32 SL systems. The auto-encoder correctly captures the general shape of the lensed source features and the deflector. The main advantage of this method is that it does not rely on any assumptions about the light profile of the lens and the source, and is therefore also able to deblend   complex lenses. This is true in particular   for UNIONS~J155923$+$314712 in which the deflector is an edge-on spiral. However, as is seen in the residual column of Fig.~\ref{Fig: deblending lens 1}, it tends not to correctly deblend some of the high-frequency features from the original images. The auto-encoder clearly distinguishes   companion galaxies from lensing features since in general, if companions are reconstructed by the auto-encoder, they appear in the deblended LRG image but not in the deblended lens source image (see Fig~\ref{Fig: deblending lens 1}). One exception is UNIONS~J113952$+$303204, where one of the companions is very close to the arc of the lensed source and is then mistaken for a lensing feature. 

In its current implementation, our method provides reliable photometry of the deflectors, according to tests performed on simulations, but performs less well on the photometry of the lensed source  because auto-encoders do not capture all high-frequency features, especially the fainter ones. This results in  a loss of flux in some of the deblended lensed sources, and indeed the residuals displayed in Fig.~\ref{Fig: deblending lens 1} show signal at the locations of small features of the lensed sources. However, we can still obtain reliable photometry of the lensed sources by subtracting the deblended deflectors from the original images and then by carrying out the photometric measurements on the subtracted images. Future work with auto-encoders will focus on better representation of the high-frequency signals contained in the data.

In conclusion, inspecting the auto-encoders results allows us to confirm the presence of promising potential lensing features in our SL sample. The deblended images support our classification for all our SL candidates. In some cases they enhance the visibility of features that are hardly visible in the original low-contrast images, and thus can provide  significant help during the visual inspection step of future lens searches.

\begin{figure*}[htbp]

    \includegraphics[width=\linewidth]{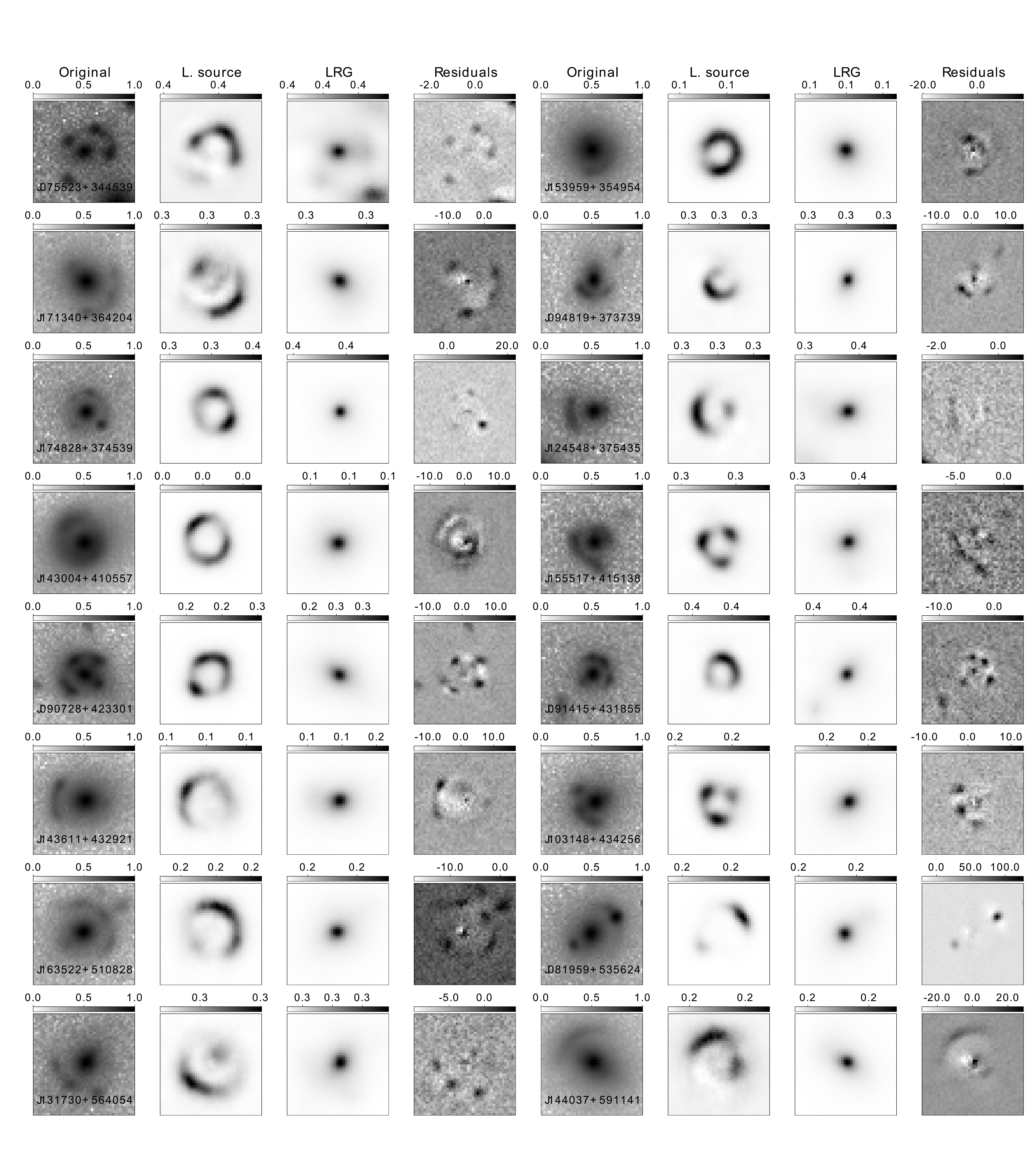}
   
    \textbf{Fig. 7.} Cont.

\end{figure*}

\section{Automated modeling of best lens candidates}
\label{modeling}

Future lens surveys will discover tens of thousands or hundreds of thousands of galaxy-scale strong lenses. From these lenses it will be necessary to define subsamples of objects suited to specific science goals. For example, not all lenses are useful for constraining dark matter substructures and not all lenses are useful for studying galaxy evolution,   constraining the initial mass function, or  inferring the lens mass-to-light ratio. It is therefore crucial to obtain a basic characterization of the lens and sources properties of the candidates already at the level of the discovery catalog. In the absence of redshift information this boils down to the Einstein radius of the system, its external shear, and   the light properties of the lens and of the source. In this section we use our  32 best objects in the SL category as a test bench for a simple automated lens characterization pipeline.

\subsection{Modeling pipeline}

We model the lens mass, lens light, and source light profiles using simple analytical profiles. We use a SIE profile to model the lens mass distribution to which we add external shear (SIE$+\gamma_{\rm ext}$ model). The light distributions of both the lens and the source are represented as single elliptical Sersic profiles. As we show below, these simple models are sufficient to fit most of our lens candidates.

The pipeline is based on the {\tt Lenstronomy}\footnote{\url{https://github.com/sibirrer/lenstronomy}} Python package \citep{Birrer2015,Birrer2018} and has two main steps, a pre-sampling optimization, followed by a full Markov chain Monte Carlo (MCMC) sampling. The pre-sampling step uses the particle swarm optimization (PSO) method \citep{Kennedy1995}, which ensures that we initialize the MCMC with model parameter values close to the maxima of the posterior probability distribution. We then perform the MCMC sampling using the {\tt emcee}\footnote{\url{https://github.com/dfm/emcee}} package, which is a Python implementation of the affine-invariant Markov chain Monte Carlo ensemble sampler \citep{Goodman2010,emcee}. 
While lens modeling is a fairly easy task for isolated lenses, it is complicated by the presence of intervening objects unrelated to the lens, which introduce spurious light contamination. Such objects should be masked to avoid being mistakenly identified as lensed images of the source. This masking procedure is addressed in different ways by different authors. \citet{Shajib2020} modeled $23$ lenses from the SLACS sample \citep{Auger2009}, and specifically chose systems that do not contain any contaminating sources of light. \citet{Night2018} did not restrict their sample, but adopted a circular mask with a fixed radius of 3.9\arcsec, which selects only the regions of the data dominated by the lensed source light. 

\begin{figure}
   \centering
   \includegraphics[width=\linewidth]{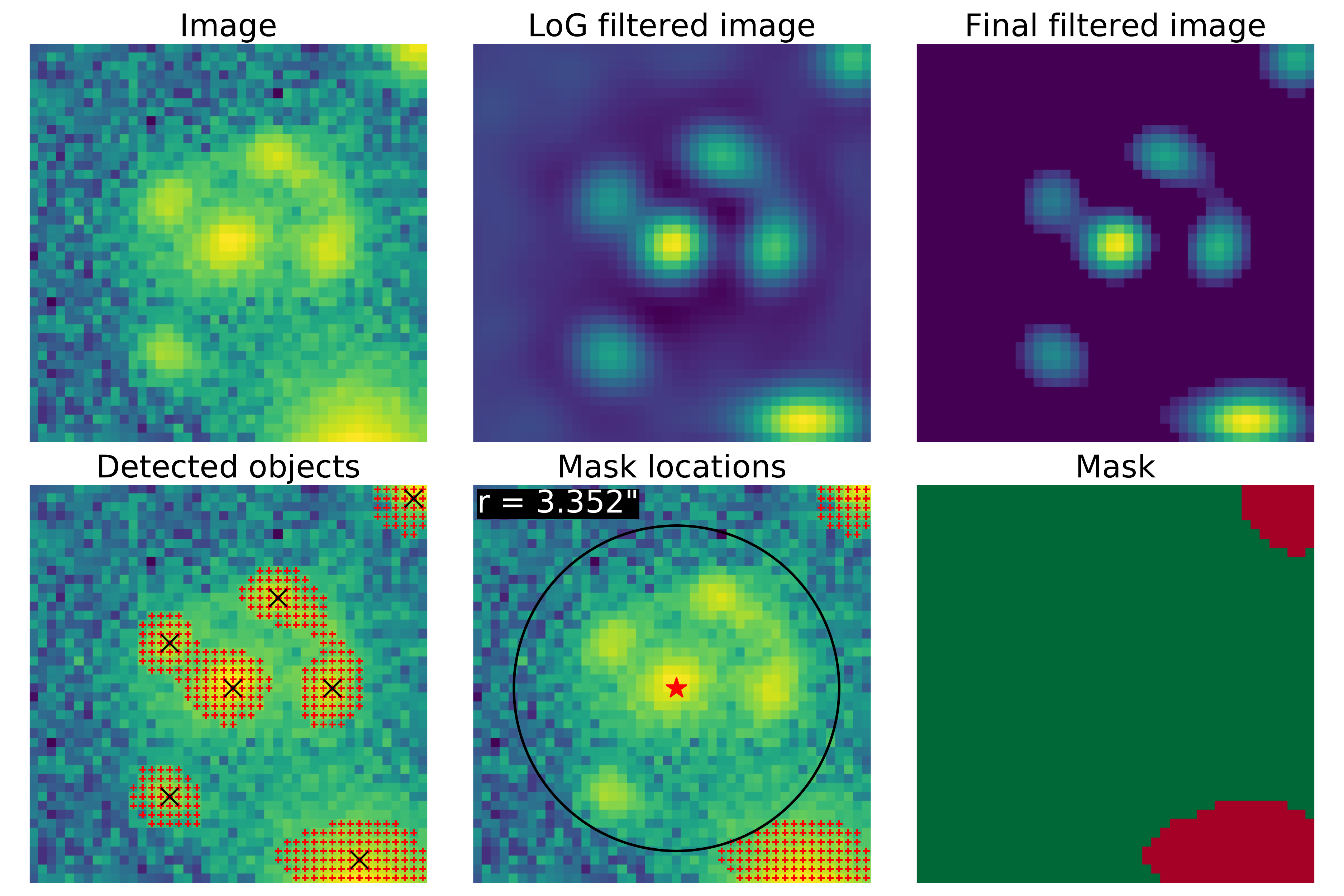}
      \caption{Illustration of our masking procedure. The upper row shows the original CFIS image in the $r$ band, followed by the image after applying the LoG filter in the top middle panel. This image is then thresholded and all pixels below 6$\sigma_{\rm sky}$ are set to zero, as shown in the top right panel. The bottom row shows the detected peaks, whose centroids are marked with black crosses. The bottom middle panel displays the estimated lens size as a black circle and the red areas indicate what we consider as contaminants. These are masked as the red area on the bottom right image. }
         \label{Fig: Masking}
   \end{figure}

In our case the masking algorithm is designed to adapt to systems with very different angular sizes. Figure~\ref{Fig: Masking} depicts the steps of the following algorithm:

\begin{enumerate}
    \item A Laplacian of Gaussian (LoG) filter is applied to the image to highlight areas of the image with a strong flux gradient. 
    \item All pixels whose flux is below a threshold of 6$\sigma_{\rm sky}$, where $\sigma_{\rm sky}$ is the {\rm rms} background noise, are set to zero. 
    \item All of the nonzero pixels in the filtered image are located, and   the locations of the peaks are identified. Peaks are defined as local maxima in the image     detected using the peak\_local\_max function of the {\tt skimage} Python package. We require that detected maxima be separated by more than one pixel from each other in order to be considered a peak.
    \item The detected objects   near the  center of the image are assumed to be the lensing galaxy and lensed images and/or arcs from the source light. These objects are used to estimate the angular size of the lens--source system. This is done by sorting the list of detected peaks by their distance from the image center:   the first peak corresponds to the lens galaxy--LRG, and    the second detected object is one of the images of the lensed source. In order to not mask part of the light from the lensed source image, the lens size is estimated to be eight pixels larger than the distance from the center to the second detected object.  All of the brightest pixels farther from the center are treated as contaminant light to be masked. The mask itself is created by using the subset of all nonzero valued pixels from step 3 whose location is farther from the center than the lens system size (red cross-hatched  areas outside the black circle in Fig.~\ref{Fig: Masking}).
    \item  The final mask is a Boolean array of the same shape as the original image, containing zeros for all pixels that are to be ignored in the modeling and a value of one elsewhere. At each pixel marked with a red plus sign (+) in Fig.~\ref{Fig: Masking}, the surrounding pixels within a circular area with a radius of two pixels are set to zero. The final mask is shown in the last frame of Fig.~\ref{Fig: Masking}. For the majority of the  32 SLs we modeled, a visual inspection of the robustness of the automated masked convinced us that it always masked out all the spurious neighboring light  that would have been deleted by hand on an object-by-object basis. Only very minor corrections would have been applied (see discussion  in Sect.~\ref{Subsec:modeling results}).
\end{enumerate} 

Before  applying {\tt lenstronomy} to our 32 SLs, we adopt realistic priors on the different parameters. First, for the mass and light profiles of the lens, we constrain the axis ratios between the semi-minor and semi-major axes, $q = b/a$, using Gaussian priors centered on a value of $q=0.8$, with a standard deviation of $\sigma=0.1$. This choice was motivated by the results presented in \citet{Kelvin2012}, where  over 100\,000 galaxies of the Galaxy And Mass Assembly(GAMA) survey were modeled, finding distributions in eccentricities  peaking at $(1 - q)\simeq0.2$. In addition, we expect some similarity between the ellipticity of the deflector mass and deflector light profiles; however, small deviations are allowed. We therefore apply Gaussian priors, with $\sigma=0.01$, on both of the ellipticity parameters of the lens mass ($e_1^{\rm m}$, $e_2^{\rm m}$) that are centered on the corresponding values of the light profile ($e_1{\rm l}$, $e_2^{\rm l}$). 

\begin{figure}
   %\centering
   \includegraphics[width=9cm]{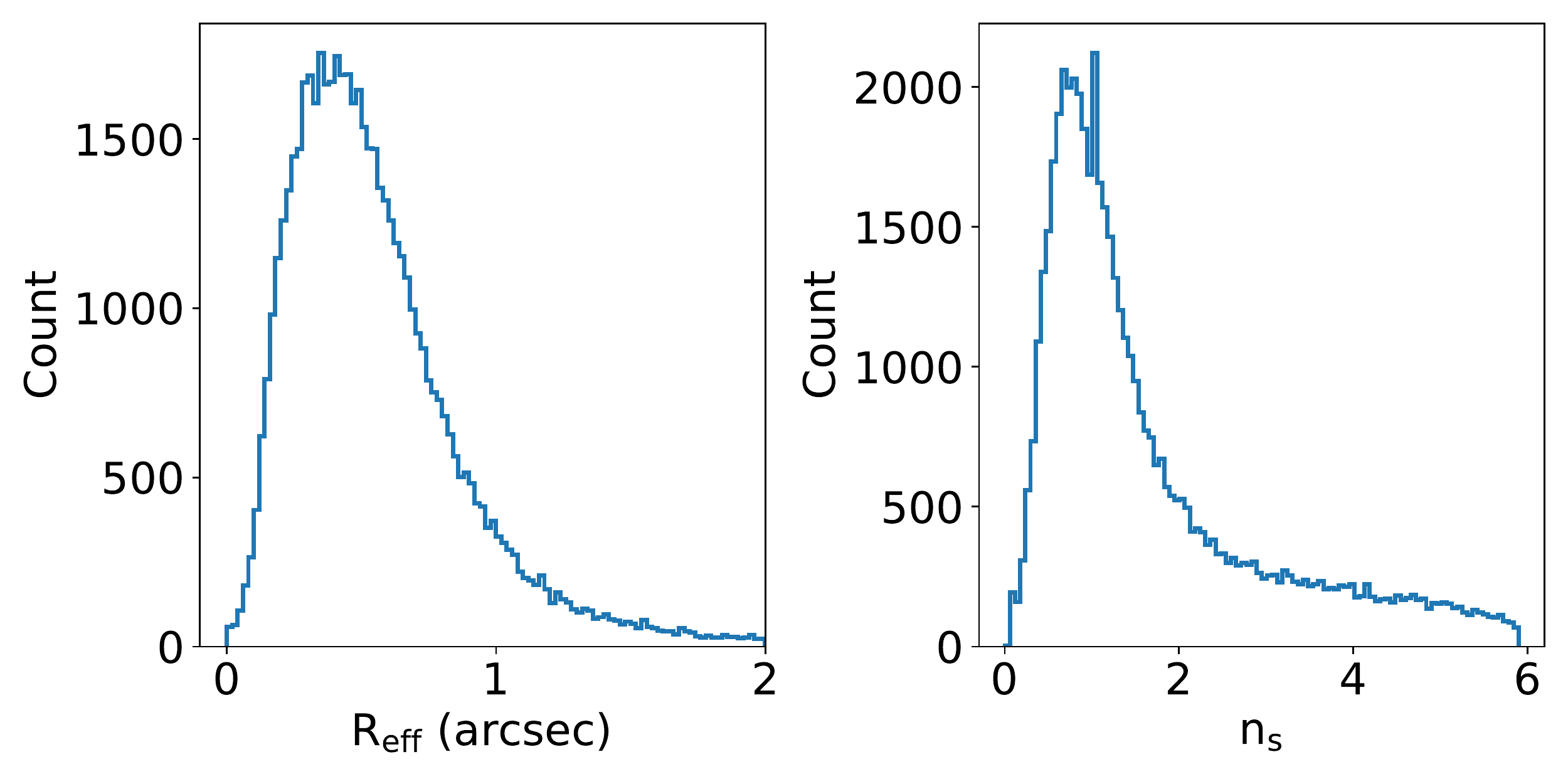}
      \caption{Priors used for the effective radius $R_{\rm eff}$ and Sersic index $n_s$ of the source light. The priors are derived from 56 062 galaxies from the COSMOS survey.}
         \label{Fig: source priors}
   \end{figure}

In addition to the priors on the ellipticity parameters, we also constrain the effective radius $R_{\rm eff}$ and Sersic index $n_{\rm s}$ of the source light through the use of prior probability distributions obtained from a catalog of  56 062 galaxies from the COSMOS survey that were modeled using a single Sersic profile to serve as a training set for the GALSIM\footnote{\url{https://github.com/GalSim-developers/GalSim}} galaxy image simulation software \citep{Rowe2015}.  We show these prior distributions in $R_{\rm eff}$ and $n_{\rm s}$ in Fig.~\ref{Fig: source priors}. For simplicity, we do not assume any covariance in the prior distribution of these two parameters nor do we assume any covariance with the deflector's flux or magnitude.

\begin{figure}
   \centering
   \includegraphics[width=\linewidth]{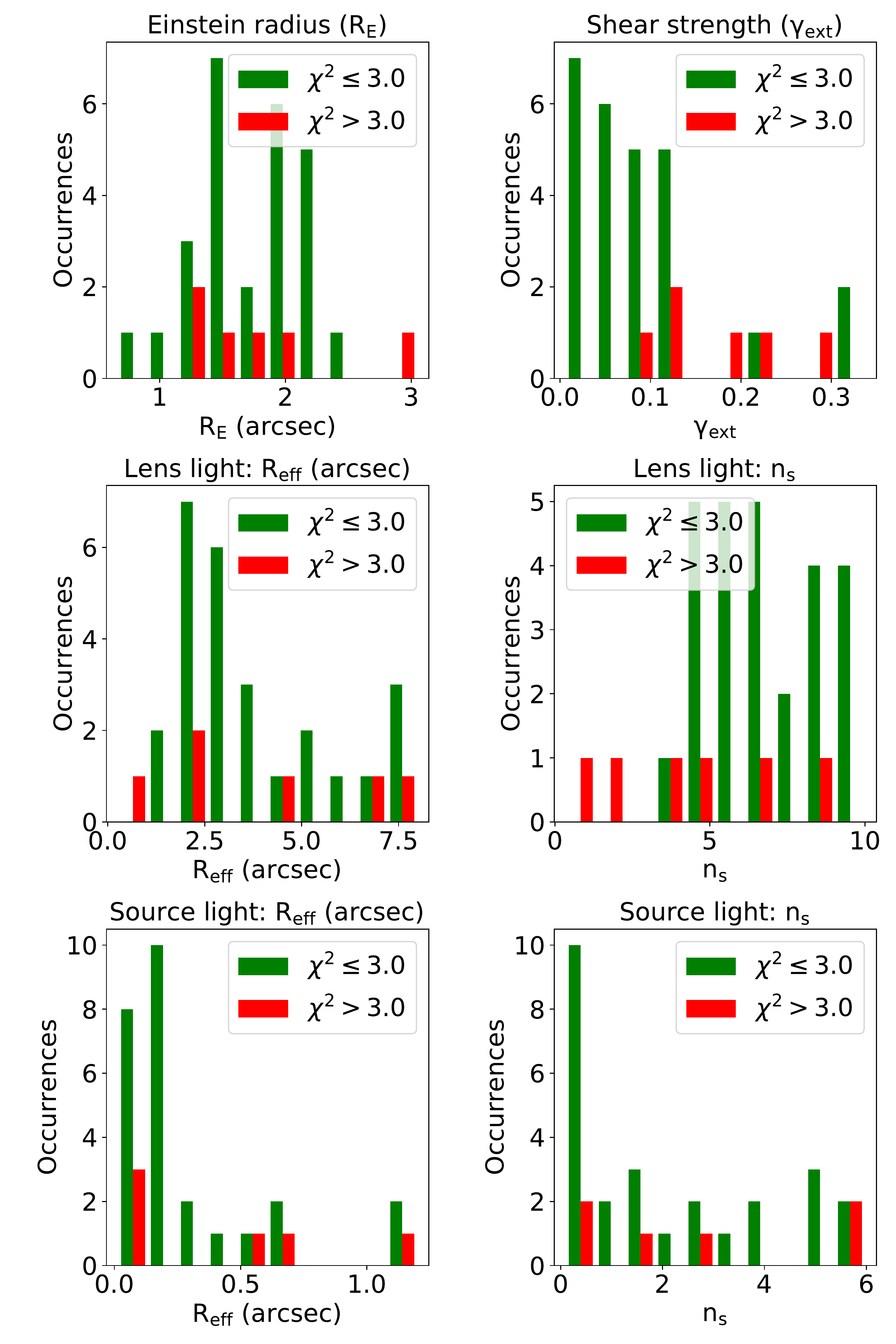}
      \caption{Parameter distributions from our lens modeling results using the SIE$+\gamma_{\rm ext}$ model. The top row gives the Einstein radii and external shear values of the mass model. The middle and bottom rows show the effective radius and Sersic index,  for the lens and source light, respectively. The models with an acceptable reduced $\chi^2$ ($\leq 3.0$) are shown in green and those with a reduced $\chi^2 > 3.0$ are shown in red (see text). After a manual fix of the masking procedure, the bad fits  improved to $\chi^2 < 3.0$.}
         \label{Fig: param histograms}
   \end{figure}
   
\subsection{Modeling results}
\label{Subsec:modeling results}

We apply our {\tt lenstronomy}-based pipeline with the priors described above to the best 32 lenses found with the CNN search. Figure~\ref{Fig: param histograms} shows histograms for the model parameters describing the deflector mass, the source light, and the deflector light profile. The Einstein radii of the lenses are in the range 1.2\arcsec $< R_{\rm E} <$ 2.5\arcsec\ and the external shear strengths are all $0.3$ or less. The CNN is biased to find lenses with Einstein radius matching the range of the training set, but the Einstein radius range 1.2\arcsec $< R_{\rm E} <$ 2.5\arcsec\ in the SL sample also highlights the fact that the visual inspection predominantly selects obvious wide-separation lenses with deblended counter images. Seven lenses have models with a shear compatible with zero, but since no account for the strong correlation of external shear and internal ellipticity is made, this must be interpreted carefully. For the lens light the effective radius and Sersic index distributions peak at $2.5\arcsec$ and $5.0$, respectively, which is not surprising as our lenses are selected among a sample of LRGs. The source galaxies are generally much smaller than the deflector LRGs, with the distribution in effective radius peaking at $R_{\rm eff}\sim 0.2\arcsec$ and with Sersic index peaking at $n_s\sim1.0$, also unsurprising given that source galaxies are often low mass and/or star-forming disks.
In Figs.~\ref{Fig: modeling mosaic 1}--\ref{Fig: modeling mosaic 7} we show mosaics of the modeling results for the 32 lenses. The Table with the lists the best-fit parameters obtained from these fits is available ate the CDS. In spite of the simplicity of our models the residuals are acceptable for most systems (i.e., with a mean reduced $\chi^2$ close to 1.0).
%(see Table~\ref{tab: modeling results table}).
However, setting a limit of 3 on the reduced $\chi^2$ allows us to spot outliers. These objects are indicated with a red rectangle in Figs.~\ref{Fig: modeling mosaic 1}--\ref{Fig: modeling mosaic 7} and show that the bad fits are due to complex lens light profiles beyond Sersic, inaccurate masking, or objects in the source plane being modeled as if they were in the lens plane or vice versa. A notable example is UNIONS~J155923$+$314712 (first row of Fig.~\ref{Fig: modeling mosaic 3}) which has a bright elongated deflector that is not well described by a single Sersic profile. This specific example also shows the limitation of our pre-selection of LRG deflectors as obviously this system has a lens with an edge-on disk and a bulge component. In all cases with $\chi^2>3.0$, we were able to immediately identify the problem and correct it in a simple way, bringing the new $\chi^2$ value close to 1.0. % (objects indicated with an asterisk in Table~\ref{tab: modeling results table}). 

Four additional objects have $\chi^2 > 3$ (UNIONS~J113952$+$303204, UNIONS~J165710$+$315052, UNION~J075346$+$341633, UNIONS~J112053$+$342146). For three of these the mask produced by our automated procedure is simply not large enough to cover all of the light contaminants. In the case of UNIONS~J165710$+$315052, part of the lensed source light in the bottom right corner of the image is mistaken for a companion, while the contaminant directly to the right of the deflector is treated as a lensed source image. This is a very specific configuration that would be extremely difficult to accommodate in an automated masking procedure for thousands of objects, as will be the case with future wide-field surveys. Nonetheless, since we only have 32 objects, we can afford to create new customized masks, and we show in Figs.~\ref{Fig: modeling mosaic 1}--\ref{Fig: modeling mosaic 4} the results with new masks directly beneath the results with the automated masks. In each subpanel the two modeling results are enclosed in the red dashed rectangles, and the manual masks significantly improve the residuals. 

We have a total of six modeling failures. Of these six, one fails due to the deflector light being more complex than for the rest of the sample, while three are due to imperfect masking that can certainly be improved in future versions of the pipeline. One of the failures, UNIONS~J081959$+$535624, is in fact known as a lensed quasar \citep{Inada2009} and therefore fails because our modeling procedure does not allow for point sources (see Fig.~\ref{Fig: modeling mosaic 8}). This point will be addressed in future work. For the 25 successful automated models, the average modeling time is 2 hours for a 44 pixel $\times$ 44 pixel stamp, which is well achievable in the context of future surveys like \textit{Euclid}, which is expected to find 20--30 SLs per day.

\section{Discussion and conclusions}

In this paper we presented the design of an automated pipeline to find galaxy-scale strong lenses using convolutional neural networks and applied it to the CFIS wide-field optical imaging survey being carried out with the 3.6 m CFHT in Hawaii. We used only the deep and sharp $r$-band images for which the median seeing is 0.6\arcsec\ down to a 10$\sigma$ depth of $r=24.6$. We used 2\,500 $\deg^2$ of CFIS in the present work since the survey is still ongoing; it is expected to reach a total area of 5\,000 $\deg^2$ of the northern sky, when completed.  

In training our CNNs, we used data-driven simulations where the light distribution of the lens plane is taken directly from the data. This naturally includes the PSF, companions, and noise properties of the actual data and any companion galaxies. The lens mass, modeled as a SIE profile, was derived from the measured velocity dispersion of our sample of LRGs, which  we further adapted to ensure that the lensing features were systematically visible in the training set. Our background sources are   from the HST images of the COSMOS survey, and are convolved by the local CFIS PSF after lensing. 

Starting from a sample of CFIS $r$-band images of LRGs selected both from spectroscopy and color-cuts, we used a committee of three neural networks, leading to a CNN-based sample of 9\,460 objects passing a CNN score of 0.5, averaged over the three members of our committee. The adopted threshold of 0.5 on the CNN identification is based   on experiments with our validation set and on the CNN score obtained for real CFIS images of spiral galaxies identified in GalaxyZoo.

Even though the precision of the CNN committee is extremely high using a score threshold of 0.5, the large sample of two million galaxies to classify implies a large number of false positives among the 9\,460 objects passing our threshold due to the low prevalence of lenses in real data, as explained in Sect.~\ref{subsec:performance classifier}. Fortunately, visually inspecting 9\,460 objects by eye is still doable and this task was done by six independent authors of this paper. Even though  strict and homogeneous rules were set for the visual classification, we note that the six human classifiers still had very different opinions on what  a lens is  and what it is not, meaning that any automated CNN may still require a time consuming human check, for example  with citizen-science projects for future wide-field space surveys like Euclid, Rubin-LSST, or Roman. More effort should be devoted to visual inspection methods to reach better consensus between classifiers.

Following the visual inspection, we found 32 objects with striking lensing features and 101 objects that show strong signs of lensing but that need further data to confirm (i.e., higher resolution, and deeper imaging and/or spectroscopy). This represents around 0.05 candidates per square degree, which is much lower than the 1.95 and 11.95 lenses per square degree estimated in \citet{Collett2015} for LSST and  \textit{Euclid}, respectively. This number is however comparable with that found by  \citet{Canameras2020}, from which we take our lens search sample, who obtain 0.0117 lenses per square degree. The slightly larger number of lenses per square degree in our case  can be explained by the higher resolution of CFIS r-band images and the different lower limit on the Einstein radii of the simulations from the training set. 

A by-product of our simulations set is that we were able to  train auto-encoders to learn the lens light and lensed-source light separately and we then deblended the lens plane from the source plane for all 133 objects. We see this process, or future evolutions of it, as a way to infer photometric redshifts for the lens and source if many bands are available, which will be the case in future wide-field imaging surveys. With the CFIS $r$-band data alone, we still find our application of auto-encoders useful to evaluate the quality of the lens candidates, especially for the smallest Einstein radii and/or for objects with strong contrast between the lens and source light. 

Finally we developed a simple lens modeling pipeline based on the {\tt lenstronomy} software in which we adopt an SIE mass profile with external shear. We also developed an automated masking procedure to enforce only relevant objects to be modeled and avoid objects unrelated to either the lensed source or the lensing galaxy. The optimization process using particle swarm optimization followed with Markov chain Monte Carlo sampling takes on the order of two hours per object for our 44 pixel $\times$ 44 pixel stamps. With the present ground-based data, even with deep high-resolution ground-based imaging, we find that the SIE plus shear model is sufficient to fit most of the data. The main sources of failure are the following: (1) the lens light was too complex to be described by a single Sersic profile (1 object out of 32); (2) the masking procedure failed to capture the entire extent of the objects to be masked (2 objects out of 32); and (3) the misidentification of source images that should not have been masked (1 object out of 32). We consider only the last case to be a real limitation to a fully automated procedure. 

 We also produced a catalog of contaminants that mimic lensing geometry (i.e., 238 mergers, 369 ring galaxies, and 961 spiral galaxies). All these contaminants are provided in electronic form as they can be useful for future lens searches in order to train CNNs against false positives.
 
We demonstrated the possibility to build an automated pipeline to find, deblend, and model lenses in future large-scale surveys, even though there are still challenges to overcome. In particular, human intervention was required at two steps of the pipeline: in the verification of the lens candidates and, to a lesser extent, in the modeling. The importance of the visual inspection step may be decreased in future versions of the pipeline by retraining the CNNs with our catalog of false positives and by combining the information obtained with the deblending and modeling with the classification score provided by the committee. Since the CFIS imaging data are among the best available so far in terms of depth and seeing, our results, although subject to improvements, can be seen as an illustration of what can be achieved with in a single Rubin-LSST band built by stacking some of the best seeing epochs after about a month of data acquisition. With a spatial resolution three times better than the best CFIS images, \textit{Euclid} (and then Roman) will give us access to a larger number of small separation systems, especially with Einstein radii smaller than 1\arcsec. 
These data will also help us to decide on our less secure candidates, since space-based imaging will provide high signal-to-noise ratios for the lensing features of the lenses presented here, allowing us to test more complex mass models, and thus probe astrophysical questions like galaxy evolution and the structure of their dark matter halos.

\begin{acknowledgements}
This work is supported by the Swiss National Science Foundation (SNSF) and by the European Research Council (ERC) under the European Union's Horizon 2020 research and innovation program (COSMICLENS: grant agreement No 787886).
This work is based on data obtained as part of the Canada-France Imaging Survey, a CFHT large program of the National Research Council of Canada and the French Centre National de la Recherche Scientifique. Based on observations obtained with MegaPrime/MegaCam, a joint project of CFHT and CEA Saclay, at the Canada-France-Hawaii Telescope (CFHT) which is operated by the National Research Council (NRC) of Canada, the Institut National des Science de l'Univers (INSU) of the Centre National de la Recherche Scientifique (CNRS) of France, and the University of Hawaii. 
This work was supported in part by the Canadian Advanced Network for Astronomical Research (CANFAR) and Compute Canada facilities.
GV has received funding from the European Union’s Horizon 2020 research and innovation program under the Marie Sklodovska-Curie grant agreement No 897124.
RC, SS, and SHS thank the Max Planck Society for support through the Max Planck Research Group for SHS. This project has received funding from the European Research Council (ERC) under the European Unions Horizon 2020 research and innovation programme (LENSNOVA: grant agreement No 771776). 
RG thanks IoA and the Churchill College in Cambridge for their
hospitality and acknowledges local support from the French government.

This research has made use of the VizieR catalogue access tool, CDS, Strasbourg, France (DOI : 10.26093/cds/vizier). The original description of the VizieR service was published in \cite{vizier}.
This research has made use of the SIMBAD database, operated at CDS, Strasbourg, France \cite{simbad}.

\end{acknowledgements}
---------------------------------------------------------------

\bibliographystyle{aa}
\bibliography{paper}

%--------------------------------------------------------------------
\begin{appendix} %First appendix

\section{Visual inspection results}
\label{Sec:appendix visual inspection}
\begin{figure}[h!]
\centering
\includegraphics[width=9.2cm]{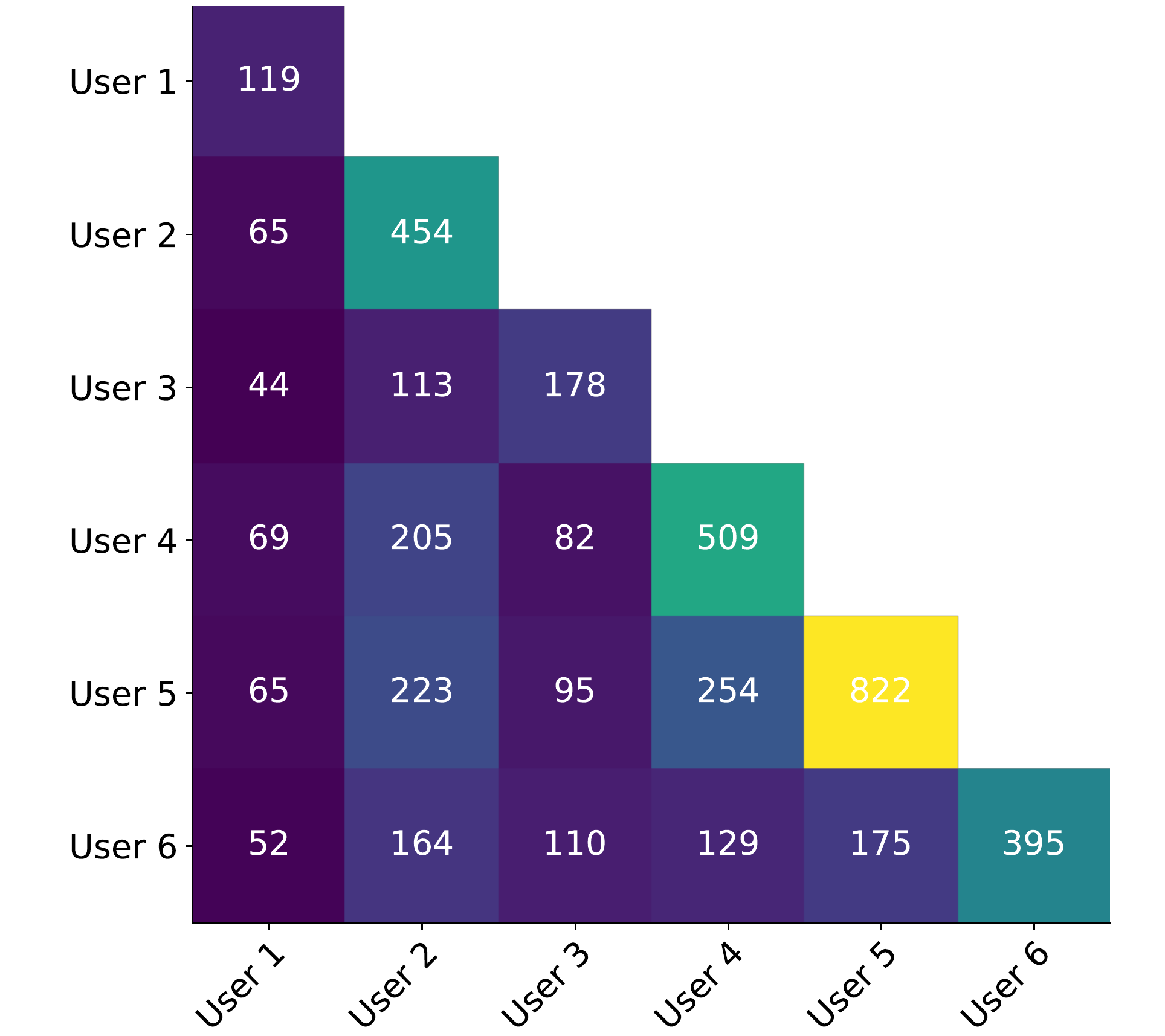}
\caption{Illustration of the overlap between the visual classification of different users. The diagonal terms show the number of SL and ML labeled by each user during the first steps of the inspection, whereas the nondiagonal terms correspond to the number of ML and SL objects that are shared between the users in the corresponding rows and columns.}
\label{Fig:stats first step}
\end{figure}

We present in Table~\ref{tab: inspection part1} the results of the single-object inspection for each user. The agreement between the human classifiers is low, as can be observed in Fig.~\ref{Fig:stats first step}, which displays the number of images labeled as either ML or SL by each user and the numbers shared between each pair of users. The discrepancy between users is less pronounced for the NL images. The category SA causes the most confusion since no image is labeled as SA by all users, and the overlap between the users selecting the greatest and lowest number of SA (User 1 and User 5) is only five objects. As a result, despite the classification guidelines, the number of ML or SL candidates varies greatly between users (see Fig.~\ref{Fig:stats first step}). Even so, the agreement between users is the greatest for the SL category, hence showing a good consensus for the objects with the most striking lensing features.

\begin{table*}[t!]
\renewcommand{\arraystretch}{1.1}
\caption{User results of the detailed visual inspection.}
\label{tab: inspection part1}
\centering
\begin{tabular}{ccccccc}\\
\hline\hline
Classification & User 1 & User 2 & User 3 & User 4 & User 5 & User 6 \\ \hline
NL             & 4492   & 4151   & 4389   & 3941   & 3398   & 4012   \\
SA              & 17     & 23     & 61     & 178    & 408    & 221    \\
ML             & 96     & 423    & 144    & 474    & 775    & 357    \\
SL              & 23     & 31     & 34     & 35     & 47     & 38     \\ \hline
\end{tabular}
\end{table*}

\section{Auto-encoder architecture}
\begin{figure*}[htbp]
   \centering
   \includegraphics[width=10cm]{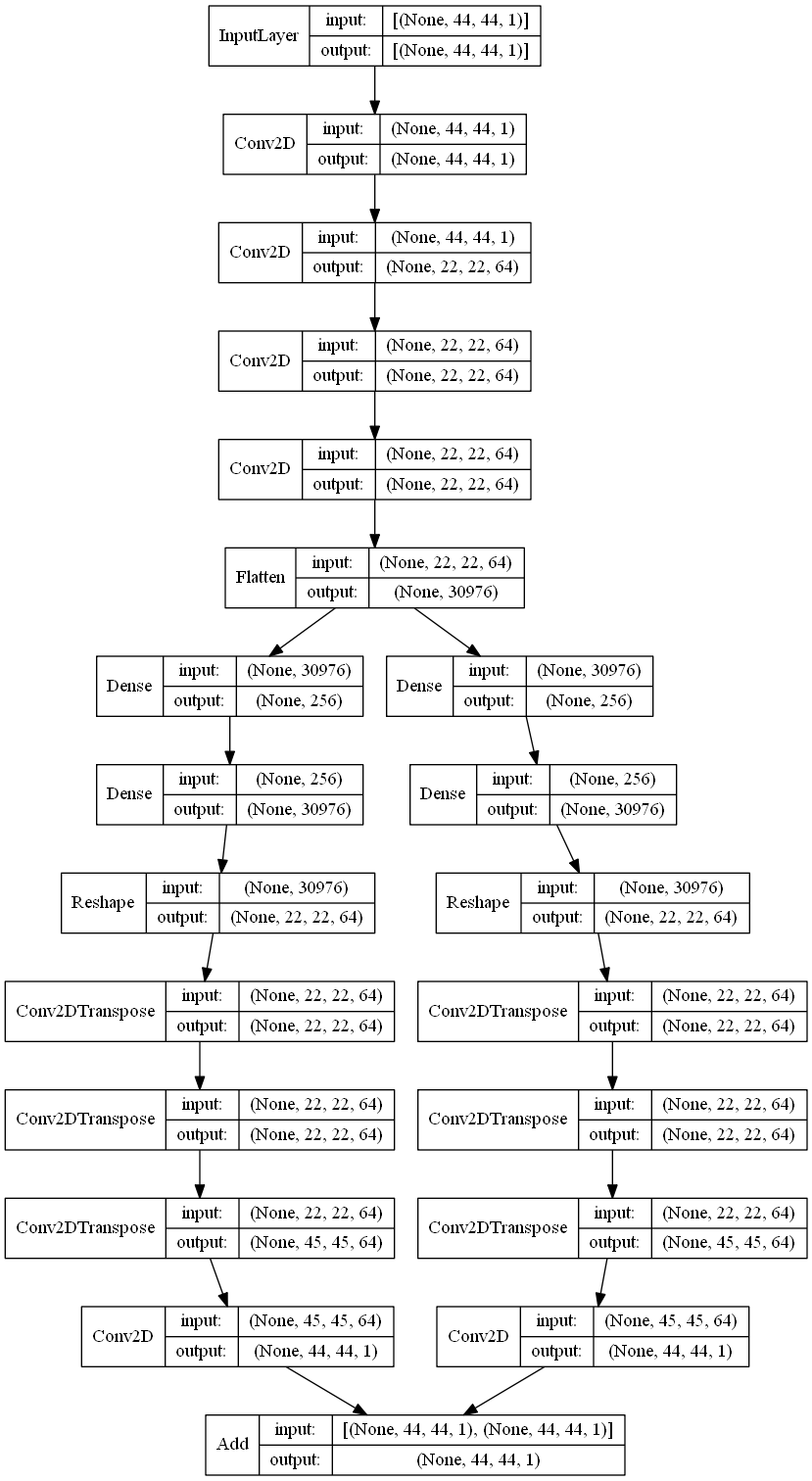}
      \caption{Architecture of the deblending auto-encoder.
Each box represents a layer and the arrows show the connections between the different layers of the networks. The names indicated in the rectangles correspond to the different layer subclasses of the Keras API used in the model. The dimensions of the input and output of each layer are indicated in brackets. For convolutional layers the last dimension corresponds to the number of filters}
         \label{Fig: auto-encoder architecture}
   \end{figure*}

We present in Fig.~\ref{Fig: auto-encoder architecture} the architecture of the deblending auto-encoder. In this scheme all layers are represented by a box and the connections between the different layers are shown with arrows. The decoder is separated in two independent and symmetrical sections that specialize in extracting the lensed source features and the deflector images. For both the encoder and the decoder we use a combination of dense and convolutional layers. The decoder part has three output layers. The last two convolutional layers output the deblended lensed-source and deflector images and the Add layer returns the sum of the two deblended images. We applied ``Relu'' activations to all neurons of the network except in the last three layers for which we used sigmoid activations in order to keep the range of the output between -1 and 1.
 \section{Maybe lens candidates} 
\begin{figure*}[p!]
   \centering
   \includegraphics[width=\linewidth]{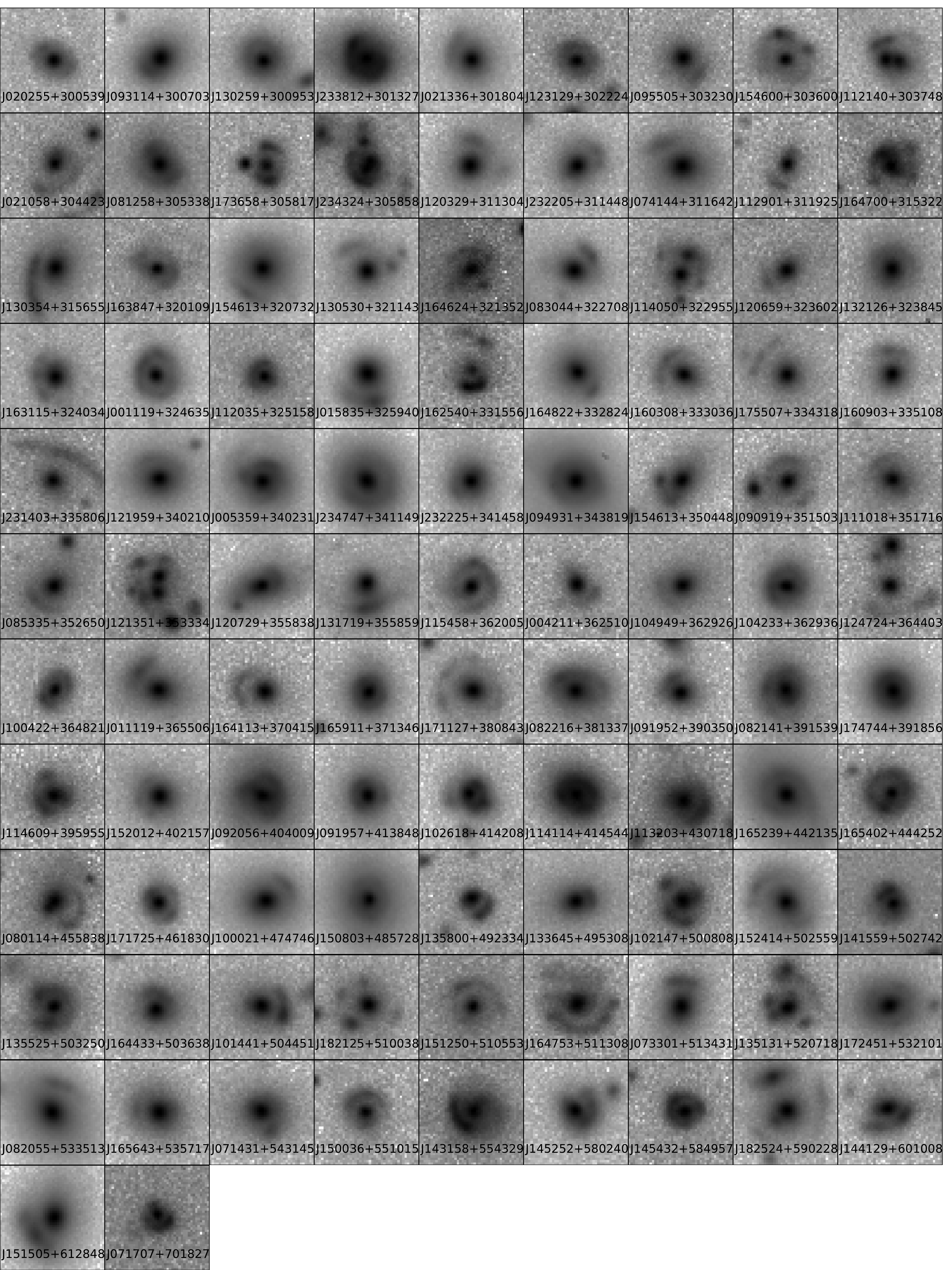}
      \caption{Images classified as maybe lenses (MLs) after the final visual inspection step.}
         \label{Fig:ML mosaic}
   \end{figure*}

 We show in Fig. ~\ref{Fig:ML mosaic} the 101 ML candidates obtained at the end of the visual inspection process. It includes all candidates displaying convincing probable lensing features but that  require follow-up observations to be confirmed.

\section{Modeling results}
%The  best-fit parameters for the 32 SL obtained after the modeling procedure are listed in Table~\ref{tab: modeling results table}.
We show in Figs.~\ref{Fig: modeling mosaic 1}--\ref{Fig: modeling mosaic 7} a mosaic with the modeling results for the 32 SL candidates.

\begin{figure*}[p!]
   \centering
   \includegraphics[width=\linewidth]{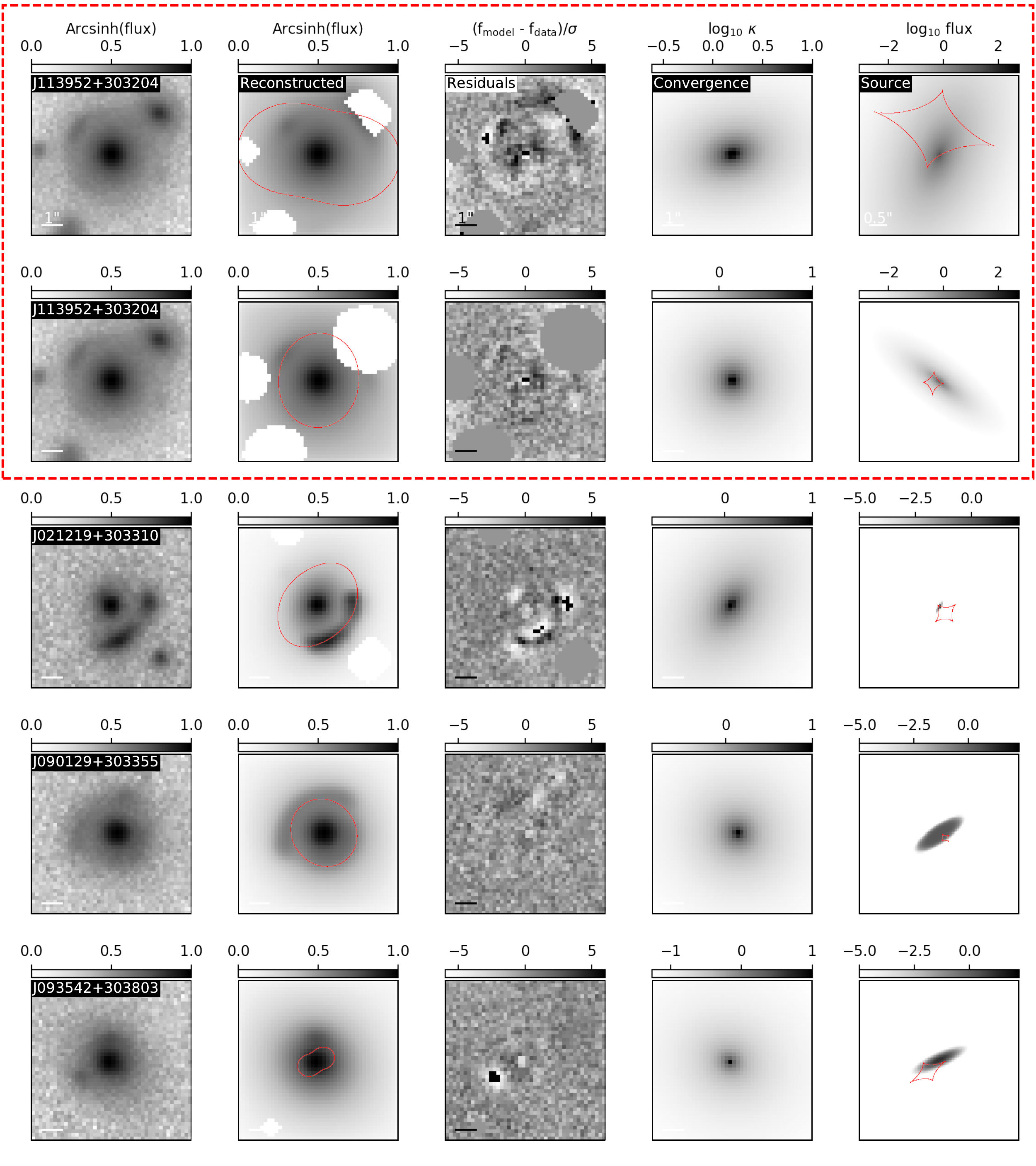}
      \caption{Modeling results for the first four of the 32 SL lens candidates. Shown inside the red dashed box are two modeling results for the same image, but  with different masks. The top row corresponds to the results from the automated masking procedure, and  in the bottom row to the results after applying a custom mask. \textit{1st column:} CFIS $r$-band image. \textit{2nd column:} Image reconstruction using  best-fit model parameters. The white regions are masked pixels corresponding to locations of neighboring objects in the observed image. In red we show the critical lines of the lens model. \textit{3rd column:} Normalized residual map of the image reconstruction. \textit{4th column:} Lens mass model convergence map. \textit{5th column:} Reconstructed source light profile (unlensed). In red are shown the caustic lines of the lens model.}
         \label{Fig: modeling mosaic 1}
   \end{figure*}

\begin{figure*}[p!]
   \centering
   \includegraphics[width=\linewidth]{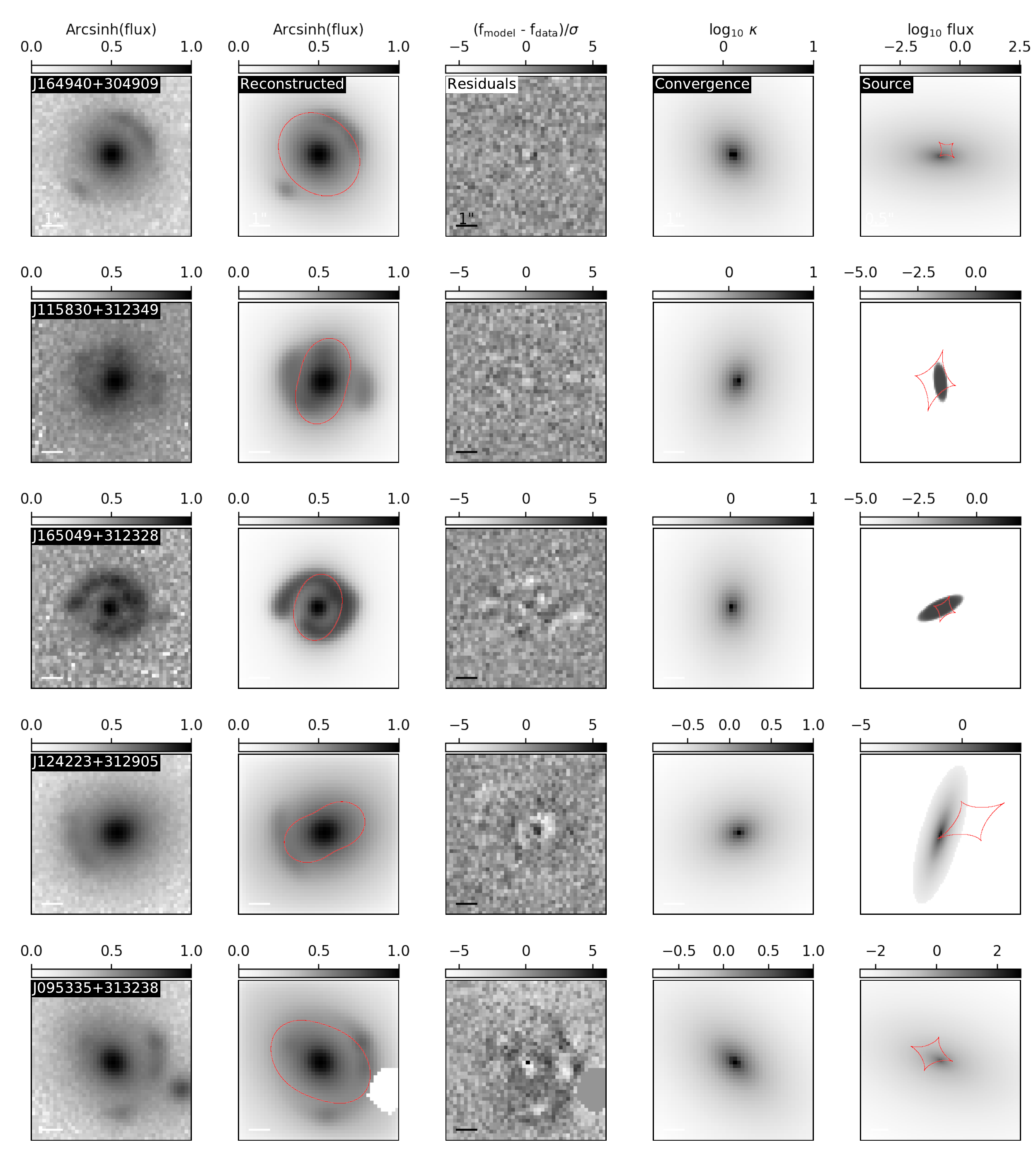}
      \caption{Continued from Fig.~\ref{Fig: modeling mosaic 1}}
         \label{Fig: modeling mosaic 2}
   \end{figure*}
   
\begin{figure*}[p!]
   \centering
   \includegraphics[width=\linewidth]{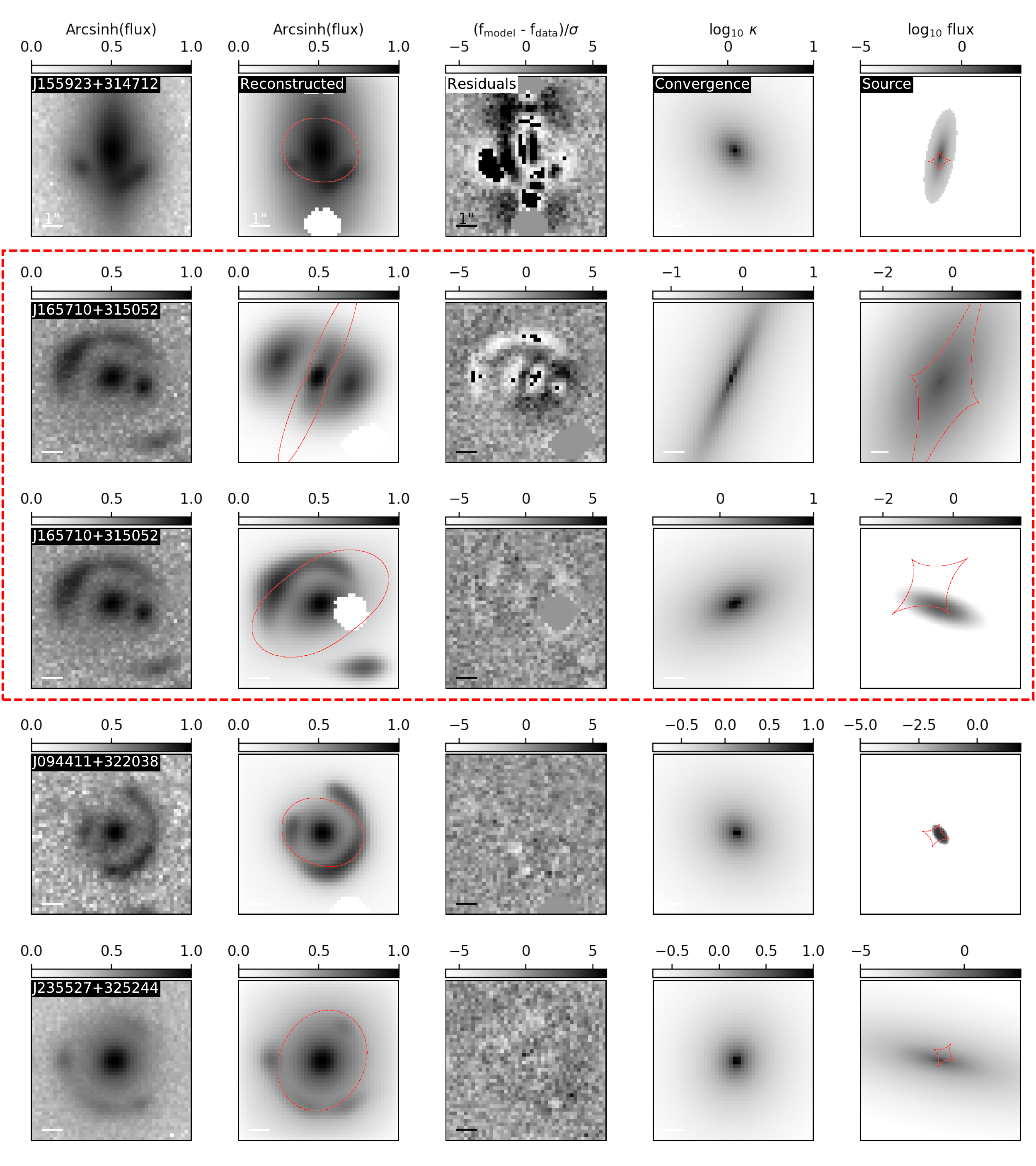}
      \caption{Continued from Fig.~\ref{Fig: modeling mosaic 2}}
         \label{Fig: modeling mosaic 3}
   \end{figure*} 
   
\begin{figure*}[p!]
   \centering
   \includegraphics[width=\linewidth]{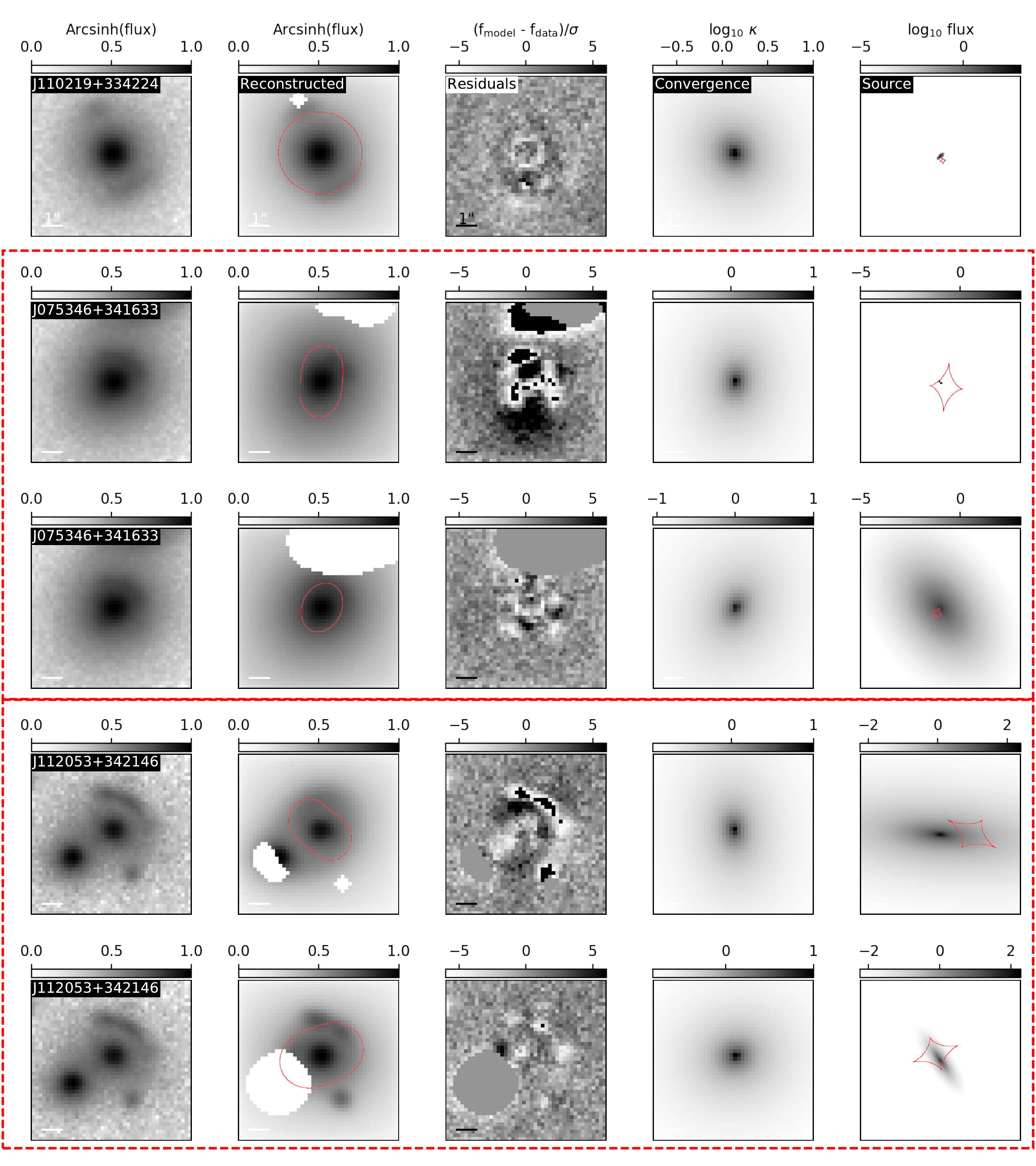}
      \caption{Continued from Fig.~\ref{Fig: modeling mosaic 3}}
         \label{Fig: modeling mosaic 4}
   \end{figure*} 
   
\begin{figure*}[p!]
   \centering
   \includegraphics[width=\linewidth]{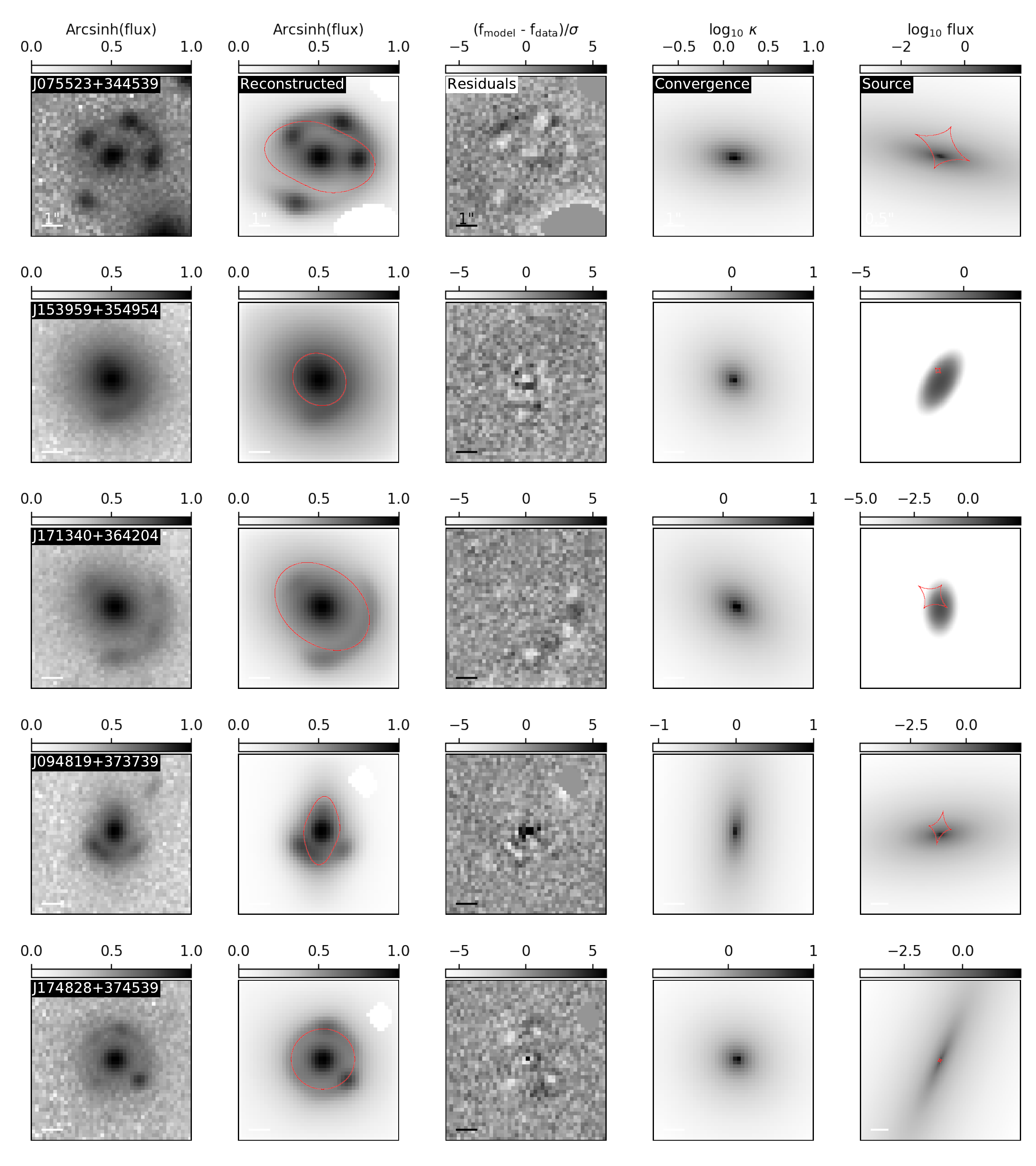}
      \caption{Continued from Fig.~\ref{Fig: modeling mosaic 4}}
         \label{Fig: modeling mosaic 5}
   \end{figure*} 
   
\begin{figure*}[p!]
   \centering
   \includegraphics[width=\linewidth]{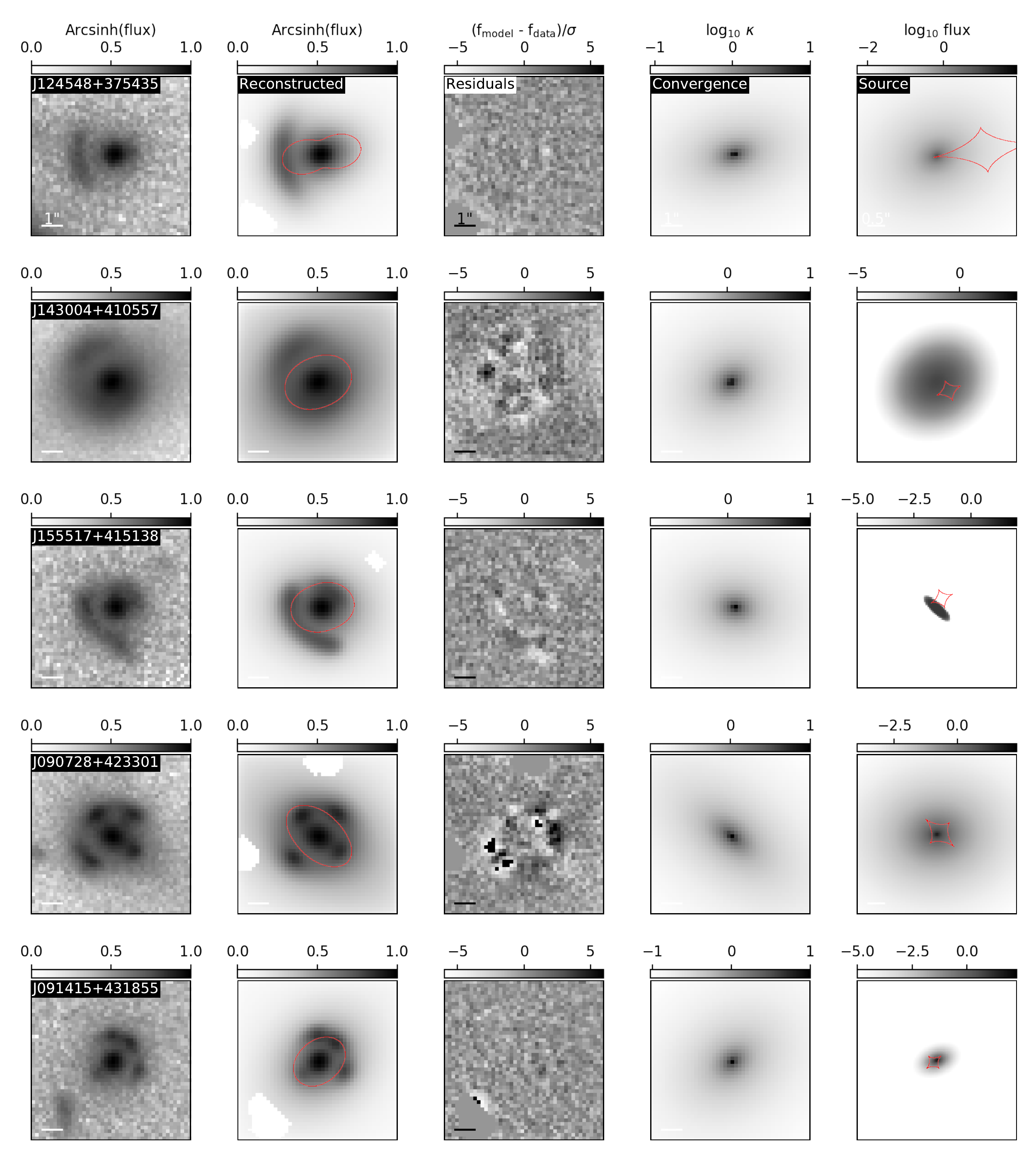}
      \caption{Continued from Fig.~\ref{Fig: modeling mosaic 5}}
         \label{Fig: modeling mosaic 6}
   \end{figure*} 
   
\begin{figure*}[p!]
   \centering
   \includegraphics[width=\linewidth]{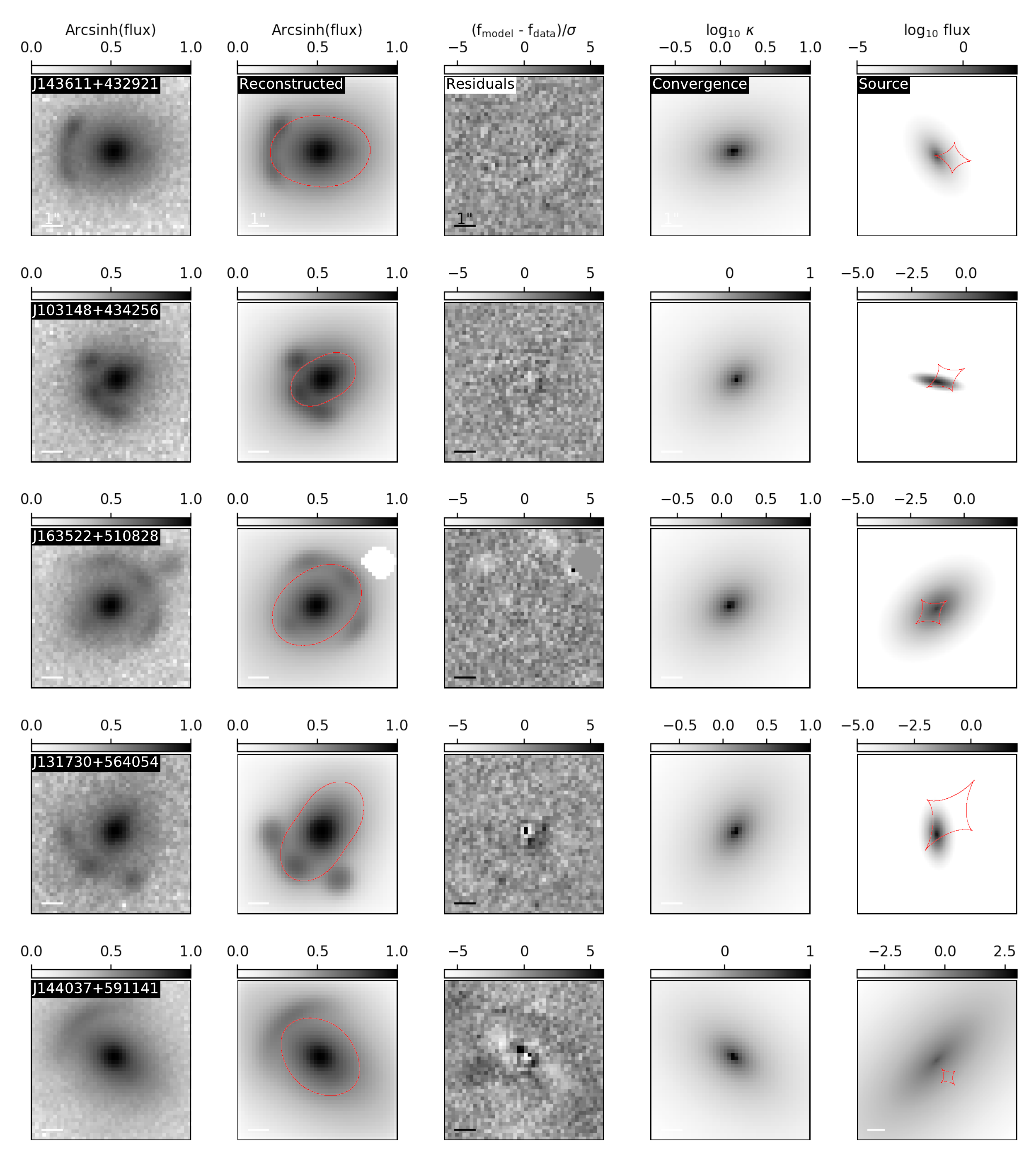}
      \caption{Continued from Fig.~\ref{Fig: modeling mosaic 6}}
         \label{Fig: modeling mosaic 7}
   \end{figure*} 

\begin{figure*}[p!]
   \centering
   \includegraphics[width=\linewidth]{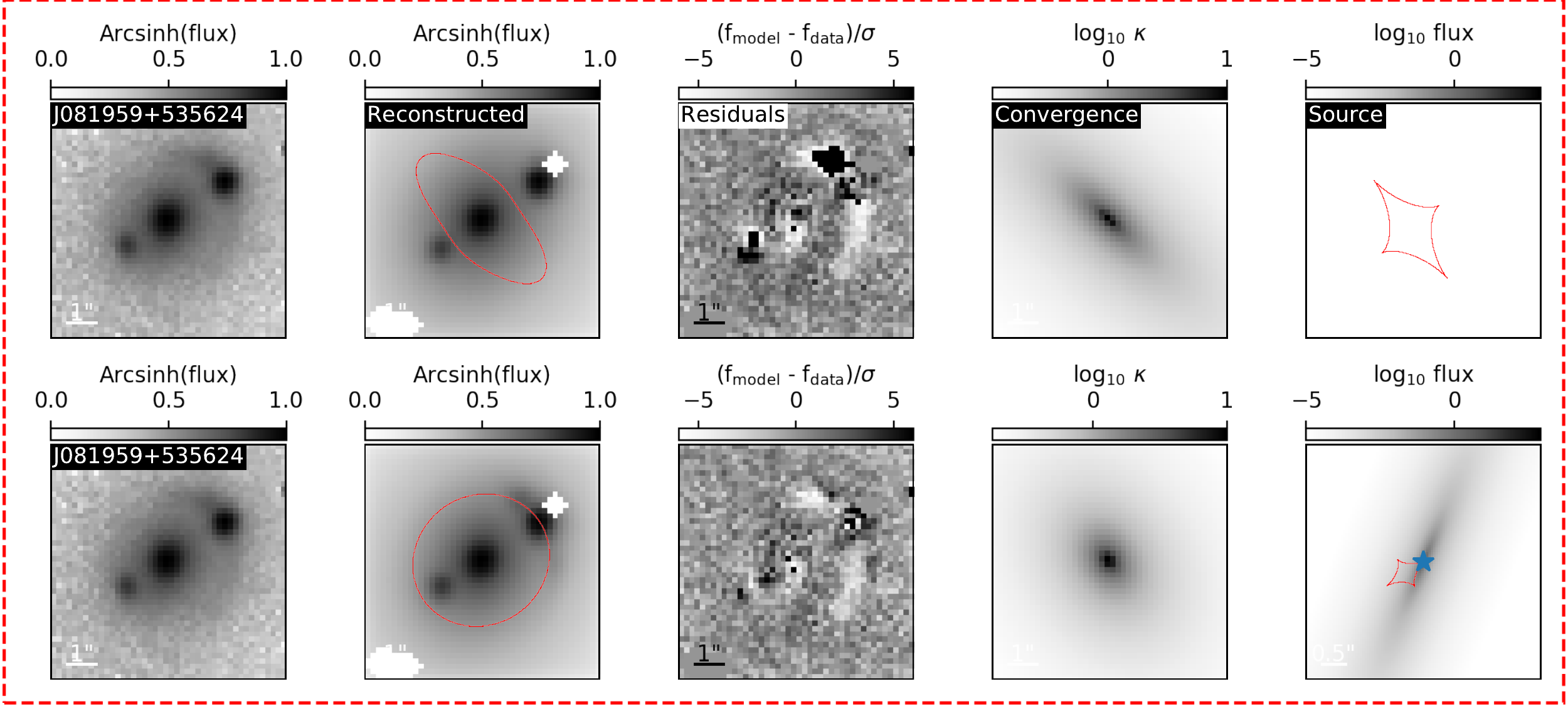}
      \caption{Continued from Fig.~\ref{Fig: modeling mosaic 7}}
         \label{Fig: modeling mosaic 8}
   \end{figure*}

\section{Examples of contaminants}
\begin{figure*}[p!]
 
   \centering
   \includegraphics[width=\linewidth]{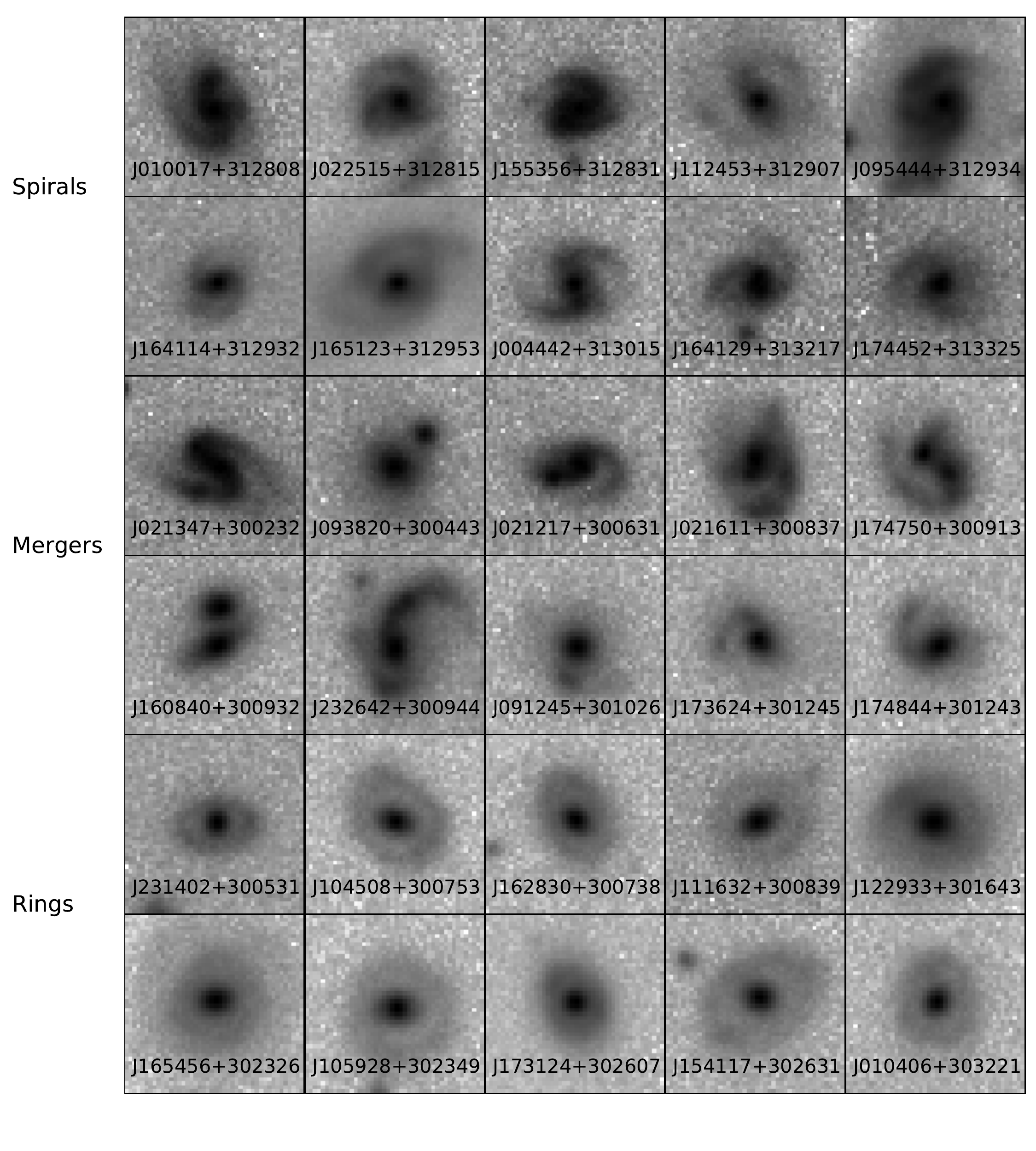}
      \caption{Examples of images classified by at least one person as spirals (first two rows), mergers (two rows in the middle), or ring galaxies (last two rows) during the visual inspection. Each of them were identified by at least one user.}
      \label{Fig: FP mosaic}
        
   \end{figure*}

We present in Fig.~\ref{Fig: FP mosaic} examples of images taken randomly from the 238 mergers and 361 ring galaxies, and the  950 identified after the visual inspection.

\end{appendix}

\end{document}